\documentclass[conference]{IEEEtran}
\IEEEoverridecommandlockouts
\usepackage{cite}
\usepackage{hyperref}

\usepackage{amsmath,amssymb,amsfonts}
\usepackage{algorithmic}
\usepackage{graphicx}
\usepackage{textcomp}
\usepackage{xcolor}
\usepackage{xspace}
\usepackage{array}
\usepackage{makecell}
\usepackage{
  tikz,
  pgfplots,
  pgfplotstable
}
\usetikzlibrary{
  shapes.geometric,
  arrows,
  pgfplots.groupplots,
  matrix
}
\usepackage{booktabs}
\usepackage{multirow}
\usepackage{graphicx}
\usepackage{subcaption}
\usepackage{listings}
\usepackage[table]{xcolor}
\usepackage[most]{tcolorbox}
\usepackage{enumitem}
\def\BibTeX{{\rm B\kern-.05em{\sc i\kern-.025em b}\kern-.08em
    T\kern-.1667em\lower.7ex\hbox{E}\kern-.125emX}}
\pgfplotsset{compat=1.18}   

\begin{document}

\newcommand{\hlt}[1]{\textcolor{blue}{#1}}
\newcommand{\todo}[1]{\textcolor{red}{[Todo: #1]}}
\newcommand{\binghui}[1]{\textcolor{brown}{[Binghui: #1]}}
\newcommand{\ziyi}[1]{\textcolor{green}{[Ziyi: #1]}}
\newcommand{\darren}[1]{\textcolor{red}{[Darren: #1]}}
\newcommand{\levi}[1]{\textcolor{purple}{[Levi: #1]}}
\newcommand{\dinil}[1]{\textcolor{red}{[dinil: #1]}}

\newcommand{\mirroredK}{\reflectbox{K}}

% method name 
\newcommand{\sol}{{\textsc{DarKnight}}\xspace}
\newcommand{\CircledGOne}{
    \tikz[baseline=(char.base)]{
        \node[shape=circle, draw=black, fill=black, inner sep=0.6pt] (char){\textcolor{white}{\scriptsize{G$_1$}}};
    }
}

\newcommand{\CircledGTwo}{
    \tikz[baseline=(char.base)]{
        \node[shape=circle, draw=black, fill=black, inner sep=0.6pt](char){\textcolor{white}{\scriptsize{G$_2$}}};
    }
}

\newcommand{\CircledGThree}{
    \tikz[baseline=(char.base)]{
        \node[shape=circle, draw=black, fill=black, inner sep=0.6pt] (char){\textcolor{white}{\scriptsize{G$_3$}}};
    }
}

\newcommand{\CircledGFour}{
    \tikz[baseline=(char.base)]{
        \node[shape=circle, draw=black, fill=black, inner sep=1pt] (char){\textcolor{white}{\scriptsize{G$_4$}}};
    }
}

\newcommand{\darkcirc}[1]{%
\tikz[baseline=(char.base)]{
  \node[circle, fill=black, inner sep=1.2pt] (char)
  {\textcolor{white}{\footnotesize #1}};
}}

\newtcolorbox{obsbox}{
  colback=gray!6,
  colframe=gray!55,
  boxrule=0.4pt,
  arc=1pt,
  left=5pt, right=5pt, top=3pt, bottom=3pt,
  boxsep=2pt,
  breakable,
  enhanced jigsaw
}

\newlist{obsitem}{itemize}{1}
\setlist[obsitem]{
  leftmargin=1.2em,
  labelsep=0.4em,
  itemsep=1pt,
  topsep=2pt,
  parsep=0pt,
  label=\textbullet
}

\title{\sol: Retrieval-Guided Agentic Framework for Breaking Dark Web CAPTCHA Gates
}
% \author{Anonymous submission}

\author{\IEEEauthorblockN{Binghui Wu, Yeo Zi Yi, Darren Tan, Manit Baser}
\IEEEauthorblockA{\textit{National University of Singapore} \\
\textit{Singapore}}
\and
\IEEEauthorblockN{Levente Csikor, Dinil Mon Divakaran}
\IEEEauthorblockA{\textit{A*STAR Institute of Advanced Intelligence and Compute} \\
\textit{Singapore}}
% \and
% \and
% \and
% \and
}

\maketitle

\begin{abstract}
Cyber Threat Intelligence (CTI) investigators who monitor dark web marketplaces for leaked credentials and breach listings must pass the diverse CAPTCHAs that operators deploy to block automated collection.
Existing solvers either need unscalable per-site fine-tuning or rely on commercial Vision-Language Model (VLM) APIs, which surrender operational sovereignty: provider policies, Tor-exit blocking, government regulation, or model updates outside the investigator's control can halt or alter the solver overnight, and per-call costs recur with every session and retry. 
We present \sol, a self-improving agentic framework designed to run on a single workstation GPU inside an investigator's Tor crawler and solve CAPTCHAs with small, locally hosted open-weight VLMs that no provider decision can withdraw, without per-market training or third-party services.
For each CAPTCHA, \sol retrieves successful and failed past attempts on similar images (dual-signal RAG), composes an image-specific preprocessing pipeline, and learns from the site's pass/fail verdict.
In a case study on four real-world dark web marketplaces, agentic tool selection outperformed parameter scaling.
A deployable open-weight VLM with \sol reached 83.2\% on the most heavily defended market, 37.2 percentage points above its base model, outperforming even the frontier Claude Opus~5 (44\%).
\sol recovered 48.2\% of baseline failures at a 7.6\% degeneration cost, indicating that selective tool use can avoid much of the degradation caused by uniform preprocessing.

\end{abstract}


\section{Introduction}
\label{sec:intro}

Dark web marketplaces demonstrate resilience against law enforcement efforts, with new platforms rapidly emerging to replace those that are disrupted~\cite{europol2026evolving}. 
The scale of illicit activity continues to grow: data breaches posted on underground forums increased by 43\% in 2024, compromised credentials rose by 34\%, and the number of ransomware group leak sites grew by 53\% over the same period~\cite{lerner2025underground}.
Beyond data and credentials, dark web forums host exploit listings that are subsequently linked to known vulnerabilities in production systems~\cite{samtani2022exploits}. Phishing kits are openly traded and deployed at industrial scale~\cite{bijmans2021phishing,phishingecosystem2026}.
The emergence of malicious LLM services, sold openly on underground markets to lower the barrier to crafting phishing lures and malware, further amplifies the threat~\cite{lin2024malla}.
Cryptocurrency facilitates pseudonymous settlement of nearly all transactions across these markets~\cite{dearden2023crypto}.

Security teams monitor dark web marketplaces to detect, as early as possible, newly listed data breaches, emerging ransomware strains, and exposed customer records before they are exploited~\cite{CTI-Eurosp-2025, europol2026evolving}.
The value of this monitoring scales with coverage and speed, which makes automated, large-scale collection a requirement rather than a convenience.
To resist such automation, marketplaces deploy several defensive layers, including headless-browser blocking, dynamic link generation, request-rate limiting, and CAPTCHAs~\cite{importanceofdarkweb-2025-IJIS, gan-2022-captcha, vibhute2025multimodal}. %, yannikos2023scraping}.
Most of these layers can be circumvented through careful crawler engineering~\cite{crator, vibhute2025multimodal}.
CAPTCHAs are the exception.
Deployed at registration, login, and checkout, they are designed to admit humans and turn away bots~\cite{yannikos2024captchas-types, audran2022ticktock}.
It is the chokepoint that if overcome cheaply and at scale, makes automated dark-web measurement feasible; otherwise collection stalls behind manual intervention~\cite{gan-2022-captcha}.

What sits at that gate is not a clear web CAPTCHA.
Mainstream services such as reCAPTCHA, hCaptcha, and Cloudflare Turnstile are largely absent from dark web sites for structural reasons~\cite{audran2022ticktock, yannikos2024captchas-types, recaptcha, 10.1145/3477142}.
They require the client to contact a third-party endpoint the site operator does not control. 
On a hidden service, that adds the risky de-anonymization layer and logging surface; neither do legit CAPTCHA provider serve the dark web market. 
Moreover, those endpoints also degrade or block Tor exit traffic.
For these anonymity reasons mainstream CAPTCHAs are not used on the dark web~\cite{audran2022ticktock}.
The visitors compound the issue: many Tor users disable JavaScript to shrink their fingerprinting surface, which makes the behavioral, slider, and interactive challenges common on the clear web impractical to deploy.
Dark web site operators therefore build their own~\cite{yannikos2024captchas-types, gan-2022-captcha, importanceofdarkweb-2025-IJIS};
the result is a population of bespoke, locally hosted, server-rendered CAPTCHAs~\cite{yannikos2024captchas-types}.
Due to the lack of JavaScript, they tend to be less diverse than commercial offerings using standard image libraries, random fonts and additive noise.
The answer is bound to a server-side session, often at low character entropy.
The dominant form is the OCR text CAPTCHA.
It appears in a few recurring variants such as distorted Latin alphanumeric strings, image-rendered arithmetic, and non-Latin script.
Non-Latin rendering is a low-cost anti-OCR measure on regionally targeted markets.
It defeats common off-the-shelf recognizers trained predominantly on Latin text.

Existing empirical evaluations of modern CAPTCHAs study the clear web systems above~\cite{wu2026mca, zhang2025capture}.
They explicitly place dark web challenges out of scope~\cite{Usenix-2023-captcha}, leaving this design space comparatively unmeasured.
% We include Chinese-script CAPTCHAs as a representative non-Latin case, since non-Latin rendering is a low-cost anti-OCR measure that defeats recognizers trained predominantly on Latin text.
Prior works approach solving dark web CAPTCHAs from two directions, neither of which fits a dark web Cyber Threat Intelligence~(CTI) deployment.
The first trains a recognizer per CAPTCHA type.
CNN (Convolutional Neural Network) and GAN (Generative Adversarial Network) solvers reach high accuracy on the styles they are trained on~\cite{yannikos2024captchas-types, gan-2022-captcha, audran2022ticktock}.
However, they require labeled data and retraining for each new style. %, which is not scalable. 
% \hlt{In fact, we  evaluated Tesseract and PaddleOCR\levi{cite this} on our dataset, and their performance fluctuates across the datasets expressing this very phenomenon (cf.\ref{app:paddleocr_and_tesseract}).}
Besides, they do not generalize across the heterogeneous, churning set of markets a collector must track.
The second is general visual solving with vision-language models (VLMs).
Currently, they can solve a range of clear web CAPTCHAs~\cite{captcha-Usenix-2025, deng2025oedipus, wu2026mca}.
On the dark web, a VLM has so far been used only as a component: an off-the-shelf model embedded in a single-market navigation tool~\cite{vibhute2025multimodal}.
There, solving is a step in data acquisition, not the object of study. 
Consequently, prior dark-web work has not characterised how VLM-based solvers behave across model scales, preprocessing pipelines, or agent designs.
These are the questions a measurement study of CAPTCHA solving must answer.
Neither direction studies dark web CAPTCHA solving as a problem in itself.
The unstudied space spans heterogeneous market styles, including the non-Latin scripts of regionally targeted markets.
It sits under the operational constraints of CTI deployment.
And it calls for a solver that adapts without per-type retraining.

Frontier commercial LLM APIs are not good options for a CTI solver, because they leave the investigator without operational AI sovereignty: control over whether, how, and at what cost the solver keeps running. 
Even a frontier model that solves every CAPTCHA today would have to be removed from the pipeline by decisions the investigator does not control. 
First, provider usage policies prohibit using the models to circumvent CAPTCHAs, and such policies can be tightened at any time. Second, regulation can cut access without warning: in June 2026, a US export-control directive led Anthropic to suspend access to two of its frontier models for all customers worldwide for about three weeks~\cite{anthropic2026fable}.
Third, providers decide which networks they serve. 
When the collection host routes its traffic through Tor, major providers challenge or block Tor exit traffic with access-denied errors and forced human-verification prompts, so reachability cannot be assumed~\cite{zhang2020torexit}. 
Moreover, Tor rotates its exit endpoint regularly; to a provider this resembles an account hopping over IP addresses, which injects additional verification mid-run and stalls an unattended pipeline until a human intervenes. 
Fourth, providers update and retire model versions on their own schedule, and solving behavior changes with them: on our most heavily defended dataset, four Claude models range from 1.4\% to 44\% accuracy (cf.~App.~\ref{app:additional_baselines}). 
Finally, the per-call cost recurs with every session and retry, since several attempts may be needed before a page opens.
Closed APIs are therefore not a practical option (nor do they perform well; see App.~\ref{app:additional_baselines}); the deployed solver must be open-weight and self-hosted, so that the investigator runs a fixed model checkpoint that no provider decision can withdraw or alter.

Two further requirements follow from the field setting.
The solver must be small enough to run on a single workstation GPU (see Sec.~\ref{sec:models} for details) for CTI deployments ($\leq 30B$ models) rather than a GPU cluster in a datacenter.
The solver has one hard timing constraint and one soft one. 
The hard constraint is server-side session expiry. 
A correct answer arriving after the session has timed out is rejected, and there is no client-side workaround~\cite{yannikos2024captchas-types}. 
The soft constraint is submission-rate detection. Markets that escalate on rapid submission can be defeated by holding the solver output and submitting after a randomised delay. 
The hard constraint binds the solver design. The soft constraint binds only the submission policy.
% And it must respect a timing window with two edges.
% If a solve runs long, the server-side session might expire and a correct answer is rejected.
% If requests come too fast, the market escalates to a harder challenge or locks the session~\cite{yannikos2024captchas-types}.
% The solver has to sit between these two limits.

% \dinil{better to rephrase this as gaps; see below}
% As we show next, meeting that profile is necessary but not sufficient.

We identify three gaps that this paper addresses.

\noindent\textbf{\CircledGOne Small open-weight VLMs for solving CAPTCHAs from the dark web within the operation constraints are understudied.}
Prior dark web CAPTCHA solvers use per-style trained models that require labelled data per market~\cite{gan-2022-captcha, audran2022ticktock}.
Other work treats solving as a downstream step in a navigation pipeline~\cite{vibhute2025multimodal}.
Modern VLM-based solvers are evaluated on the clear web, predominantly with frontier-closed or large-parameter models~\cite{captcha-Usenix-2025, deng2025oedipus}.
The capability of small open-weight VLMs in solving real dark web CAPTCHAs under time-bounded operational constraints remains understudied.

\noindent\textbf{\CircledGTwo Understudied deployable enhancement methods for open-weight VLMs.}
Prior work can be categorized broadly into two approaches: per-style retraining and frontier-API agentic solving.
% Per-scheme retraining produces strong solvers for one CAPTCHA family but does not scale across the heterogeneous, churning market landscape~\cite{wang2023experimental, khatavkar2024segfree, gan-2022-captcha}.
% On the other end, recent works use agentic frameworks~\cite{react-iclr-2023, hugginggpt-nurips-2023, octoTools} to coordinate reasoning across tool calls in CAPTCHA solving~\cite{captcha-Usenix-2025, deng2025oedipus}.
% These solvers rely on cloud APIs, which violate the deployment constraints established above.
No prior method combines agentic enhancement with a deployable open-weight backbone.

\noindent\textbf{\CircledGThree Non-Latin and heterogeneous-style CAPTCHAs from dark web are yet to be addressed.}
Existing dark web CAPTCHA work targets English-language, Western markets, and Latin-script inputs~\cite{gan-2022-captcha, audran2022ticktock, yannikos2024captchas-types}. 
Recent threat intelligence reports identify active Chinese-language underground markets as hubs that preserve operational anonymity across clearnet, darknet, and messaging platforms~\cite{crowdstrike2025apj}.
The study of any solver, classical or VLM-based, on non-Latin dark web CAPTCHAs is missing in literature.

We argue that these gaps leave us with a single viable {\em profile} to address the challenges: an open-weight solver that runs locally, stays small, and operates within a timing window. 
We hypothesize that the bottleneck is not model capacity but tool-selection strategy, informed by past solves.

To this, we design and develop \sol, a self-improving agentic solver.
\sol is given a small library of image-preprocessing tools (see Sec.~\ref{sec:framework}). 
For each CAPTCHA, \sol perceives the input and retrieves past attempts on visually similar images. 
Then, it composes a preprocessing pipeline, solves the cleaned image, and updates from the
site's verdict~\cite{react-iclr-2023}. 
The same open-weight VLM does both planning and solving, with no additional model to host. 
That closed verdict-feedback loop enables stream-level improvement without retraining.

We present this work as a case study of one framework on four marketplaces, evaluated offline; where we generalize, we show what is feasible, not what holds for every market or model.
Our primary contributions are summarized as follows:

\begin{enumerate}  

    \item We design \sol, an agentic enhancement layer for open-weight small VLMs operating under dark web deployment constraints. 
    \sol combines chain-of-thought visual analysis, a dual-signal RAG knowledge database (retains both successful and failed records of CAPTCHA solving, \S\ref{subsec:rag}), five preprocessing operations, and
    automated ground-truth feedback (\S\ref{subsec:feedback}). Across 2{,}000 CAPTCHAs from four real-world dark web marketplaces and nine model backbones, \sol improves accuracy in 30 out of 32 (93.75\%) configurations, with an average absolute improvement of 13.33\%. 
    In particular, on the most challenging dataset ($\mathcal{D}_1$), the average improvement across all backbones is +30.48\%.
    Such improvement on all four marketplaces is achieved without any per-site retraining, with a single configuration handling alphanumeric, math, and Chinese CAPTCHAs within the same mixed-style market.

    \item We present the first systematic evaluation of small open-weight VLMs (1B--30B) on real dark web CAPTCHAs under operational constraints. 
    We show that a \emph{general-purpose} 27B model equipped with \sol (83.2\%) on $\mathcal{D}_1$ outperforms the raw baselines of Step3-VL-10B~\cite{step3vl2026} (63.6\%), the strongest bare model we evaluated, a \emph{larger} general-purpose vision language model, Qwen3-VL-30B-A3B (42.4\%), and even a frontier model Claude Opus 5 (44\%), showing that adding \sol to a lightweight model beats scaling up the bare model.
    % This demonstrates that agentic tool selection enables lightweight open-source models to outperform specialized and large-scale engines that would require dedicated fine-tuning or massive computational resources.
    \item We identify and quantify the degeneration-recovery effect in CAPTCHA preprocessing. 
    Across all Planner decisions, including cases where no preprocessing is selected, \sol recovers 48.2\% of baseline failures while degrading only 7.6\% of correct predictions, yielding a $5.8\times$ recovery-to-degeneration ratio.
    The Planner selects no preprocessing in 86.9\% of cases on $\mathcal{D}_2$ and only
    26.6\% on the heavily defended $\mathcal{D}_1$, indicating that it intervenes selectively rather than uniformly.

\end{enumerate}

To support reproducibility, we publicly release the complete source code, experiment configurations, and all four annotated CAPTCHA datasets\footnote{\url{https://github.com/Binghui99/DarKnight}}.

\section{Related Work}
\label{sec:related}

\subsection{Dark Web Analytics and Anti-Crawling Mechanisms}
Dark web marketplaces continuously deploy robust anti-crawling defenses to shield illicit trade operations from automated CTI monitoring~\cite{georgoulias2021qualitative}. 
Unlike the clear web, where operators frequently deploy sophisticated, behaviorally driven validation modules (e.g., reCAPTCHA or hCaptcha), dark web administrators operate under strict Tor network constraints that fundamentally preclude third-party JavaScript execution to prevent user de-anonymization~\cite{yannikos2024captchas-types, audran2022ticktock, recaptcha, 10.1145/3477142}. 
Consequently, marketplaces rely predominantly on bespoke, locally hosted, and server-rendered CAPTCHAs. These mechanisms are characterized by aggressive background noise, hyper-kerning, variable character lengths, and mathematical puzzles intentionally engineered to defeat conventional scraping tools~\cite{importanceofdarkweb-2025-IJIS}.

\subsection{Classical ML-Based CAPTCHA Solvers}
Early automated CAPTCHA resolution relied on specialized computer vision and deep learning pipelines~\cite{bursztein2010good, gao2016simple}. 
Even prior to the emergence of multimodal foundation models, researchers demonstrated that machine learning systems could successfully attack visual reasoning CAPTCHAs that were explicitly designed to resist automated solving~\cite{gao2021visualreasoning}. 
For dark web CTI, frameworks like DW-GAN~\cite{gan-2022-captcha} and ``Tick Tock Break the Clock''~\cite{audran2022ticktock} achieve high success against bespoke challenges. However, they are operationally brittle, as they overfit to specific data distributions and require labor-intensive manual retraining whenever marketplace administrators update their visual obfuscation logic.
We do not compare against them on accuracy. Given a labeled training set for a target style these solvers score highly, but that assumption is exactly what fails in CTI: a collector tracking churning, heterogeneous markets has no per-style labels at the moment of solving and cannot retrain as styles change. 
\sol targets this regime, achieving deployable accuracy without per-style supervision.

\subsection{Generalized VLM Solvers and Agentic Frameworks}
Recent works have demonstrated the capability of VLMs in zero-shot multimodal reasoning. Recent benchmark suites such as MCA-Bench~\cite{wu2026mca} and CAPTURE~\cite{zhang2025capture} systematically evaluate the effectiveness of modern VLMs against diverse CAPTCHA schemes. 
MCA-Bench~\cite{wu2026mca} demonstrates that fine-tuned VLMs can consistently overcome cross-modal verification challenges, while CAPTURE~\cite{zhang2025capture} provides a dedicated evaluation framework for assessing LVLM performance on CAPTCHA-solving tasks. 
More recently, Cognition provides a large-scale evaluation of multimodal LLM CAPTCHA-solving capabilities and investigates defensive strategies against emerging MLLM-based attackers~\cite{cognition2025}. 
State-of-the-art agentic frameworks have further operationalized these capabilities against complex security mechanisms. 
Halligan is an agentic VLM framework, which reduces intricate visual CAPTCHAs into heuristic search problems overlaid onto graphical user interface action spaces~\cite{captcha-Usenix-2025}. 
Concurrently, the Oedipus framework utilizes an LLM-enhanced reasoning pipeline to parse and bypass advanced logic-based CAPTCHAs using a specialized Domain Specific Language~\cite{deng2025oedipus}. 
Despite their adaptability, these frontier agentic solvers uniformly rely on commercial cloud APIs for inference. 
This reliance violates the strict operational requirements of dark web CTI deployments discussed above. 
As with the trained solvers above, we therefore cannot run these frontier-API agents under our deployment constraints, and exclude them from our evaluation.
% Cloud providers proactively throttle and aggressively block IP traffic originating from known Tor exit nodes, rendering API-dependent agents entirely unstable for continuous hidden service monitoring.

\section{Threat Model}
\label{sec:threat}

\begin{figure}
    \centering
    \includegraphics[width=0.8\linewidth]{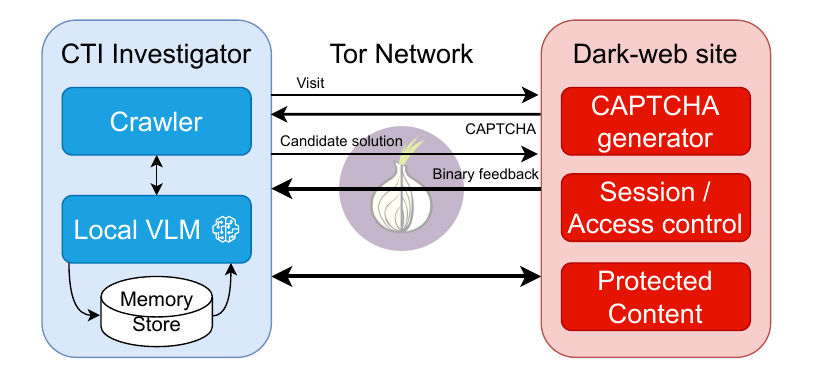}
    \caption{Threat model}
    \label{fig:threat_model}
\end{figure}

\noindent\textbf{Setting.}
We consider a security investigator who has to automate CAPTCHA solving on dark web marketplaces for scalable CTI gathering (see Fig.~\ref{fig:threat_model}).
The CAPTCHA's purpose is to gate automated access; the investigator's goal is to bypass that gate at scale.
We study the defender (the site's CAPTCHA system) by playing the attacker (the automated solver), for the larger social good. 
From the marketplace operator's perspective, the investigator is a bot.
We discuss ethical considerations in Appendix~\ref{sec:ethical}. 

\noindent\textbf{Investigator.}
The investigator operates an automated crawler over Tor.
The CAPTCHA solver interacts with the target site through a standard browser session.
It receives CAPTCHA images as rendered in the browser and submits candidate text solutions.
Each submission yields binary feedback: the protected page loads (solution accepted) or a new CAPTCHA is served (solution rejected).
The solver has no view of the CAPTCHA generation algorithm, character set, font library, or distortion parameters.
No labelled dataset of marketplace-specific image and solution pairs is available at the start.
The solver may, however, accumulate experience over time.
It may record candidate solutions, outcomes, and preprocessing strategies that succeeded or failed on visually similar images.
However, this memory is feedback-derived, not a labelled training set.
Per Sec.~\ref{sec:intro}, the solver must be open-weight, locally hosted, and resource-bounded.

\noindent\textbf{Site operator.}
The site operator controls the full CAPTCHA stack: generation algorithm, fonts, distortion, script choice, length, and presentation.
The operator also controls the server-side access-control policy.
The operator enforces request-frequency thresholds.
Exceeding a threshold triggers escalating countermeasures.
These include serving harder CAPTCHAs, invalidating sessions, and blocking the current Tor exit IP~\cite{zhang2020torexit}.
Exit-IP blocking forces the client to rotate circuits, since circuit control belongs to the client.
The operator may also update CAPTCHA designs over time in response to observed bypassing attacks.

\noindent\textbf{Scope and assumptions.}
This work targets \emph{text-based} CAPTCHAs, the dominant variant on dark-web marketplaces~\cite{gan-2022-captcha, yannikos2024captchas-types}.
The text-based scope spans Latin alphanumeric, arithmetic, and non-Latin (e.g., Chinese) variants observed in our data.
For this current work, we exclude slider puzzles, image-selection grids, audio CAPTCHAs, and any challenge requiring JavaScript execution beyond text rendering.
We assume the Tor Browser renders CAPTCHA images faithfully; i.e., what the solver receives matches what a human user would see.
We assume the operator does not adversarially adapt CAPTCHA generation to our specific system.
Tor network availability and anonymity are out of scope.
A solve is successful when the candidate solution is accepted and the protected page loads.
All experimental results are obtained offline against a static labeled dataset (Section~\ref{sec:datasets}).
Live-deployment characteristics that depend on operator adaptation are not exercised in this evaluation.

\section{Design Rationale}
\label{sec:design}
Below we present the design principles behind \sol (Sec.~\ref{sec:framework}).

\noindent\textbf{\darkcirc{1} From Monolithic Solvers to Decomposable Modules.}
A single VLM invocation that simultaneously analyzes a CAPTCHA to perform multiple operations (e.g., removing background noise and OCR), treats solving as a monolithic task. 
A CAPTCHA-solving framework should therefore decompose the task into distinct perception, planning, and execution stages, allowing each component to focus on a well-defined objective. This modular design improves controllability over individual components and enables targeted error diagnosis. This principle motivates the three-phase perceive-plan-execute architecture described in Sec.~\ref{sec:framework}.

% \noindent\textbf{\darkcirc{2} AI for perception, classical engineering for operation.} 
% Once perception identifies the noise type, the operation is well-defined image processing. 
% We apply deterministic preprocessing (e.g., denoise, distortion recognition) drawn from the established image-processing literature. 
% A cleaner image leaves the VLM with less to fight. 
% The same model is more reliable when its perceptual load is lighter.
% % Within VLM use itself, we route by specialisation: an OCR specialist handles text recognition, a generalist handles ambiguous perception. 
% % Both failure modes from prior work, specialists mis-handling math and generalists mis-handling OCR, arise from forcing one model to do work outside its strength.

\noindent\textbf{\darkcirc{2} Utilizing Transformation Experts.}
Beyond staging the pipeline, the question is who performs each step.
Identifying visual characteristics is a perception task suited to VLMs; the resulting image transformation, by contrast, is typically deterministic. 
% Once these characteristics are recognized, however, the appropriate image transformation is typically deterministic. 
VLM should be used to analyze visual distortions and interference patterns, while preprocessing operations such as denoising, line removal, and binarization should be delegated to classical computer vision techniques. This principle motivates the separation between the Planner's visual analysis component (Sec.~\ref{subsec:planner}) and the deterministic preprocessing toolbox (Sec.~\ref{subsec:executor}).

% \noindent\textbf{\darkcirc{3} Anything programmable becomes a tool, not a prompt.} 
% The same logic applies after the VLM, not just before it. 
% A VLM strong at OCR can still mis-compute simple arithmetic. 
% Once OCR extracts ``$7 + 5$'', the arithmetic is a few lines of code with deterministic correctness. 
% Asking the VLM to compute it introduces stochastic failure for a closed-form problem. 
% The same applies to regex extraction, equation evaluation, length checks, and character
% normalisation; all of these leave the model\levi{not sure if we have these outside of the VLM}

\noindent\textbf{\darkcirc{3} Deterministic Tasks are Programmable.}
% Not every CAPTCHA-solving subtask requires model reasoning. 
Separately from image preprocessing, the symbolic steps after recognition are also deterministic.
Once text is extracted, many downstream operations such as evaluating an arithmetic expression can be solved {\em programmatically} rather than by the model. 
% For example, after OCR extracts an arithmetic expression, its evaluation can be performed reliably through code. 
Programmable tasks thus should be delegated to dedicated tools, reserving the VLM for perception and interpretation while avoiding unnecessary sources of stochastic error.
This principle motivates the Executor's use of a dedicated arithmetic solver described in Sec.~\ref{subsec:executor}.

% \noindent\textbf{\darkcirc{4} Reuse past success via memory.} 
% CAPTCHAs from the same market share visual characteristics. 
% When the system has solved a similar image before, we retrieve the tool chain that worked. 
% This reduces VLM call count and stabilises behaviour across the runtime.
% The similarity metric (Sec.~\ref{subsec:???}) and memory store (Sec.~\ref{subsec:???}) instantiate this principle; they are not implementation conveniences.

\noindent\textbf{\darkcirc{4} Learn from Experience.}
CAPTCHAs from the same marketplace often share recurring visual characteristics, so a preprocessing strategy that succeeds or fails on one is likely to behave the same on similar ones. 
% causing certain preprocessing strategies to consistently succeed or fail. 
A solver should therefore retain both successful and unsuccessful attempts, and retrieve visually similar records when handling a new CAPTCHA, reusing what worked and avoiding what degraded performance.   
% When processing a new CAPTCHA, retrieving visually similar records to reuse proven preprocessing pipelines while avoiding strategies that previously degraded performance may enable continual adaptation.
This principle motivates the dual-signal RAG knowledge base (Sec.~\ref{subsec:rag}).

% \noindent\textbf{\darkcirc{5} Tools grow, weights stay frozen.} 
% When a new CAPTCHA style appears, we add a tool to the registry. 
% We do not retrain a model. 
% CAPTCHA styles drift faster than any retraining cycle an open-weight system could sustain. 
% The framework grows with the threat; the weights do not. This is the design choice that converts enhancement-over-scaling into an operational stance.

\noindent\textbf{\darkcirc{5} Plug-and-Play for Scalability.} 
Dark web CAPTCHA styles evolve continuously, making repeated model retraining impractical for operational deployments. The preprocessing capabilities should comprise modular tools rather than learned model behaviors. 
The framework is designed so that a new tool can be added without modifying the underlying model weights; we leave empirical validation of adding new tools to future work. 
This plug-and-play design may facilitate rapid adaptation to emerging defenses, and enable scaling without incurring the cost of continual retraining.
This principle motivates the modular toolbox and feedback-driven adaptation mechanisms described in Sec.~\ref{subsec:executor} and \ref{subsec:feedback}.

\section{The \sol framework}
\label{sec:framework}
% \levi{fix british/american english spelling}
% \todo{add definition of what is the "agentic" capability?}

\subsection{System Overview}
\label{sec:overview}

\begin{figure}[htbp]
\centering
\includegraphics[width=\linewidth]{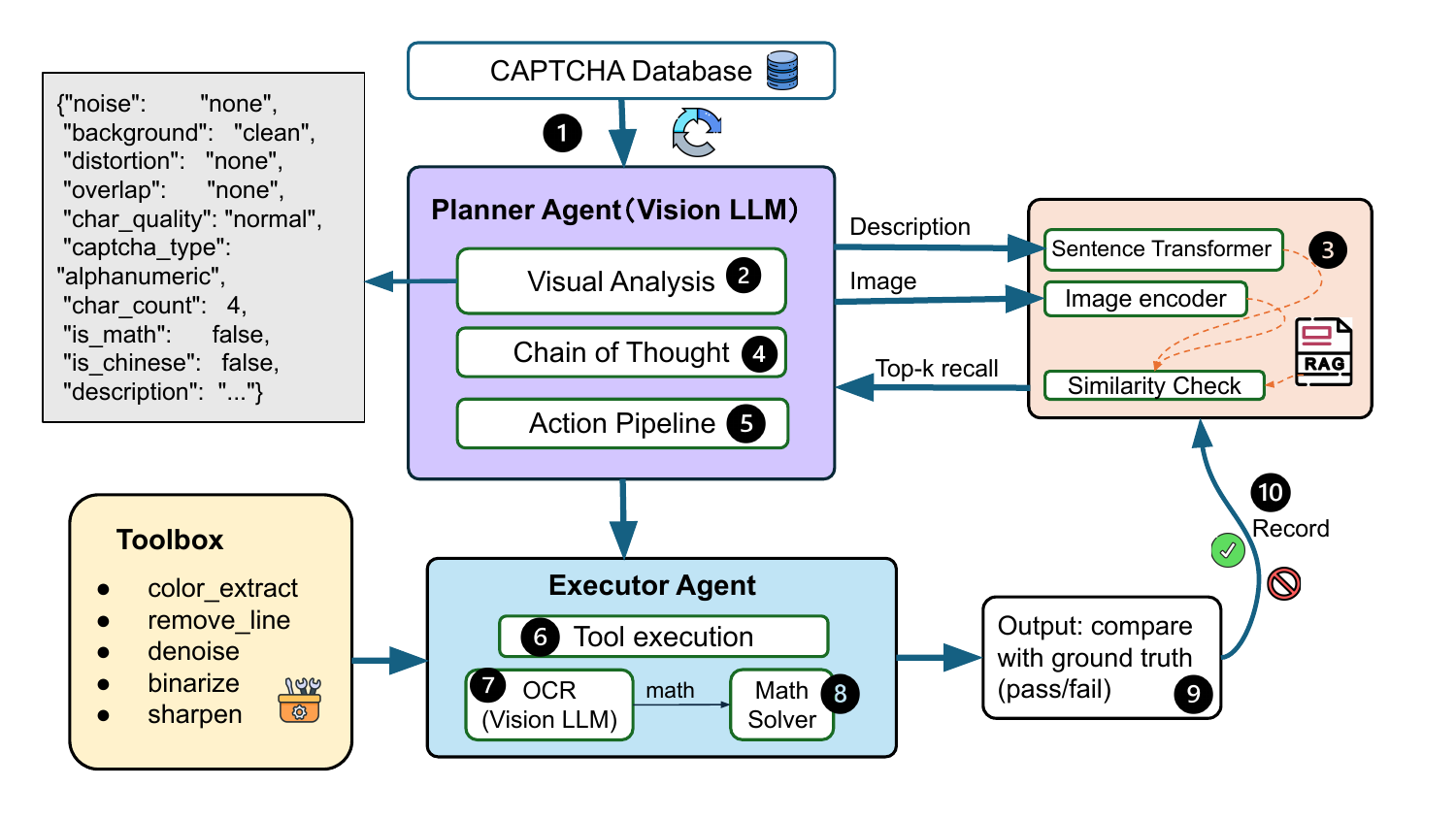}
\caption{Architecture Diagram of the pipeline}
\label{fig:architecture}
\end{figure}

Fig.~\ref{fig:architecture} illustrates the architecture of \sol.
Given a CAPTCHA image from a dark web marketplace, the system operates in a three-phase loop: \emph{plan}, \emph{execute}, and \emph{record}.

i)~In the \textbf{plan} phase, the Planner Agent receives the raw CAPTCHA image and produces a structured visual analysis describing the image's properties. 
This analysis is used to query the RAG Knowledge Base, which retrieves the most relevant past solving records (both successes and failures) based on visual and semantic similarity. The planner agent then uses the retrieved records alongside its own visual analysis and composes a preprocessing pipeline tailored to the current image.
ii)~In the \textbf{execute} phase, 
The Executor runs this pipeline step-by-step from the toolbox, passes the processed image to the OCR engine and submits the recognized text (or the solution of an equation) to the target website.
iii)~In the \textbf{record} phase, the website returns a binary pass/fail verdict.
The complete attempt, including image, visual analysis, pipeline, OCR output, and outcome, is written back to the RAG Knowledge Base regardless of the result, closing the feedback loop.
% \binghui{I move to the introduction part}
% \darren{We refer to this design as \emph{agentic} because no human engineer decides which preprocessing tools to apply.
% Instead, the Planner autonomously perceives the visual properties of each CAPTCHA, reasons about which tools are appropriate given those properties and past experience, selects and sequences the tools, and acts on that decision, all without manual intervention.
% This stands in contrast to conventional pipelines where preprocessing is fixed by design: the same sequence of operations is applied to every image regardless of its content.
% In \sol, the tool-selection policy emerges from the model's own reasoning at inference time, making the system adaptive by construction.}

As the system encounters more CAPTCHAs from the same marketplace, the RAG store accumulates site-specific experience.
This enables the Planner to progressively make better tool-selection decisions based on the past experience, \emph{with no model retraining or fine-tuning}.
This is the key design property that distinguishes \sol from prior per-style trained solvers.
A critical design choice is that the Planner VLM serves \emph{dual roles}: it performs the visual analysis and planning in the perceive phase, and also serves as the OCR engine in the execute phase.
Since the solver is already loaded, routing planning through it adds no memory overhead.
This keeps the system lightweight since only a single model is needed to be loaded, even if VLM invocation happens twice, resulting in the same memory requirements as a naive single VLM execution pipeline. 
% In practice the full agentic loop, including both VLM invocations, completes within the operational latency budget ($\sim6s$~\cite{importanceofdarkweb-2025-IJIS})  while, as shown later, improving accuracy significantly(Sec.~\ref{sec:inference-time}).
The full agentic loop introduces additional local inference time overhead. Section~\ref{sec:inference-time} quantifies this computational cost across the evaluated VLMs.

% \dinil{why do we talk about h/w here? move, probably the rest of this paragraph, to experiment section}

\subsection{Planner Agent}
\label{subsec:planner}
 
The Planner Agent is a lightweight, open-weight VLM that operates in three stages to perform visual analysis, experience retrieval, and pipeline planning.

\noindent\textbf{Stage 1: Visual Analysis.}
The Planner receives a raw CAPTCHA image and outputs a structured JSON describing eight visual properties shown in Table~\ref{tab:features}.
This structured output serves two purposes: i)~it informs the RAG retrieval query (Stage~2) and ii)~it provides the Planner with an explicit representation of the image's properties for reasoning in Stage~3.

\begin{table}[ht]
\centering
\caption{Visual features extracted by Stage~1 visual analysis.}
\label{tab:features}
\scriptsize
\renewcommand{\arraystretch}{1.1}
\begin{tabular}{lp{6cm}}
\toprule
\textbf{Features} & \textbf{Values} \\
\midrule
\texttt{captcha\_type}
  & \texttt{alphanumeric} $|$ \texttt{math} $|$
    \texttt{chinese} \\
\texttt{char\_count}
  & Integer count of CAPTCHA characters \\
\texttt{noise}
  & \texttt{none}
  $|$ \texttt{color\_lines}: multiple coloured wavy/diagonal lines of a different hue from the text $|$
    \texttt{strikethrough}: one thin dark
    line crossing all characters $|$
    \texttt{arc}: dark curved or arc-shaped
    bands $|$
    \texttt{vertical\_lines}: thin dark
    vertical strokes through characters $|$
    \texttt{dots}: scattered isolated dot
    clusters $|$
    \texttt{salt\_pepper}: fine
    pixel-level random grain \\
\texttt{background}
  & \texttt{clean} $|$ \texttt{gradient} $|$
    \texttt{noisy} $|$ \texttt{patterned} \\
\texttt{distortion}
  & \texttt{none} $|$ \texttt{wave} $|$
    \texttt{blur} $|$ \texttt{perspective} \\
\texttt{char\_quality}
  & \texttt{normal} $|$ \texttt{hollow} $|$
    \texttt{faded} $|$ \texttt{thin} $|$
    \texttt{broken} $|$ \texttt{thick} \\
\texttt{is\_math}, 
  & Boolean flag if  \texttt{captcha\_type} is math \\
\texttt{is\_chinese}
  & Boolean flag if \texttt{captcha\_type} is chinese\\
\texttt{description}
  & Free-text sentence for RAG
    text-embedding similarity \\
\bottomrule
\end{tabular}
\end{table}

\noindent\textbf{Stage 2: Experience Retrieval.}
Using the structured analysis from Stage~1, the system queries the RAG Knowledge Base (Sec.~\ref{subsec:rag}) for past  solving records on visually similar CAPTCHAs.
The query uses two signals: text similarity of the image description and perceptual hash\cite{zauner2010phash} similarity of the image itself.
The top-$k$ records, annotated with their preprocessing pipelines and website pass/fail outcomes, are appended to the Planner's context for Stage~3.
This stage is automated and does not require additional VLM inference.
 
\noindent\textbf{Stage 3: Pipeline Planning.}
The Planner receives its own visual analysis (Stage~1) and the retrieved records (Stage~2) as context, and produces a complete processing plan.
The plan specifies two components:
(i)~\texttt{preprocess\_pipeline}, an ordered list of tools to apply from the toolbox (Sec.~\ref{subsec:executor}), each with its parameters (e.g., \texttt{binarize} with \texttt{method=otsu}); and
(ii)~\texttt{post\_ocr}, a flag indicating whether to activate the math expression solver after OCR.
The Planner is prompted with a system instruction that defines the available tools, their expected inputs and effects, and the output JSON format.
When retrieved records are available, the Planner is instructed to prefer strategies that previously succeeded on similar images and avoid strategies that previously failed.
Table~\ref{tab:calls} (Appendix.~\ref{sec:apipara}) summarizes the API parameters for VLM calls.

\subsection{Dual-signal RAG Knowledge Base}
\label{subsec:rag}

The optimal preprocessing strategy for a given CAPTCHA cannot be deduced from static visual analysis alone.
Hence, \sol maintains a persistent dual-signal RAG knowledge base of past execution records, which the Planner queries dynamically at inference time. 
Such RAG knowledge base retains \emph{both} successful and failed solving attempts in distinct partitions (Table~\ref{tab:slots}), bounded to 100 records per experimental configuration. 
Successful records function as positive blueprints, guiding the Planner to replicate proven pipelines when encountering matching visual features. 
Failed records serve as negative constraints, explicitly warning the Planner against repeating historically destructive tool sequences.
This dual-signal retrieval design directly addresses the degeneration-recovery trade-off (Sec.~\ref{sec:degen-recov}). 
Rather than blindly applying an operation and discovering at test time that it degrades a specific CAPTCHA archetype, the Planner actively retrieves prior failure evidence to proactively prune hazardous tools from its execution plan.
In all experiments, the RAG store applies a cold-start protocol. 
It begins empty and is populated incrementally as each dataset is processed in a single sequential pass: the first image is solved with no retrieved experience, and each subsequent image benefits from the records accumulated by all preceding attempts.

\subsubsection{Storage Layout}

The RAGStore adopts a file-per-combination storage scheme: one JSON file is created for each unique [dataset, planner\_model, ocr\_model] triplet, stored under \texttt{rag\_store/} with the naming convention:

\[
\texttt{\{dataset\}\_\_\{planner\}\_\_\{ocr\}.json}
\]
For the primary per-dataset experiments, this storage scheme ensures isolation between different configurations, and no cross-dataset retrieval is performed. 
Each file holds up to \textbf{200 records}, distributed across two independent slots as shown in Table~\ref{tab:slots}: up to 100 successful and 100 failed records. 
This design allows us to measure the effect of \sol independently on each marketplace. 
In Sec.~\ref{sec:cross-site}, we additionally evaluate a mixed-site setting in which CAPTCHAs from all four marketplaces are processed sequentially using a single shared RAG store for each evaluated backbone.

% This ensures complete isolation between experimental configurations; no cross-querying occurs between different model or dataset combinations.

\begin{table}[h]
\centering
\caption{RAGStore slot structure.}
\label{tab:slots}
\renewcommand{\arraystretch}{1.3}
\begin{tabular}{llll}
\toprule
\textbf{Slot} & \textbf{Content} & \textbf{Capacity} & \textbf{Eviction Policy} \\
\midrule
\texttt{correct} & Successful predictions & 100 records & Least-retrieved first \\
\texttt{failed}  & Failed predictions     & 100 records & Least-retrieved first \\
\bottomrule
\end{tabular}
\end{table}

\subsubsection{Quality-Based Eviction}

Each record tracks a \texttt{\_retrieval\_count} field that is incremented each time the record is surfaced to the planner. 
When a slot reaches capacity, the system applies a least frequently used (LFU) eviction policy, displacing the record with the lowest cumulative retrieval count~\cite{king1971lfu}.
This policy is self-reinforcing as records that frequently match new queries survive indefinitely, while records that are never retrieved and thus never contribute to planning, are replaced by more relevant newcomers. 
The net effect is a memory that organically specializes toward the most frequently encountered CAPTCHA configurations.

\subsubsection{Deduplication}

On every \texttt{add()} call, the system computes the weighted similarity between the incoming record and all existing records in the target slot. If any existing record has a similarity score $\geq 0.99$, the insert is suppressed. This prevents near-duplicate examples from crowding out diverse records and preserves the breadth of the memory.

\subsubsection{Similarity Metric}

The composite similarity score combines two signals:
\begin{equation}
  s = 0.7 \cdot \cos(\mathbf{f}_q, \mathbf{f}_r) + 0.3 \cdot \cos(\mathbf{e}_q, \mathbf{e}_r)
  \label{eq:similarity}
\end{equation}

where $\mathbf{f} \in \{0,1\}^{192}$ is the perceptual hash (pHash-RGB) of the image and $\mathbf{e} \in \mathbb{R}^{384}$ is the sentence-transformer embedding of the free-text description. The pHash is computed over the RGB channels of the image (pHash-RGB)\cite{zauner2010phash}, which captures 
structural and color gradient patterns without storing the full image.
The free-text description is embedded using \texttt{all-MiniLM-L6-v2}\cite{wang2020minilm, reimers2019sbert}, a lightweight 
sentence-transformer that maps visual descriptions to a dense semantic space. Feature cosine 
(weight $0.7$) dominates because it is deterministic and encodes the exact visual properties that 
drive tool selection. Text cosine (weight $0.3$) contributes semantic context: two CAPTCHAs 
described as ``wavy blue lines over Chinese characters on a gradient background'' may share only 
partial feature overlap but should still trigger similar preprocessing decisions. Records whose 
weighted similarity falls below $\tau = 0.7$ (\texttt{RAG\_SIMILARITY\_THRESHOLD}) are dropped 
before being returned to the planner.

\subsubsection{Split Retrieval}
The \texttt{query\_split(description, features, top\_k)} method searches the \texttt{correct} 
and \texttt{failed} slots \textit{independently} and returns the top $k/2$ records from each. 
We set $k{=}5$ throughout all experiments, retrieving up to two successful and two failed records per query (with one additional record from whichever slot has the higher top match).
This guarantees balanced exposure to positive and negative examples regardless of the relative 
sizes of the two slots. Each retrieved record is presented to the planner with the following fields:
\begin{itemize}
    \item \textbf{similarity\_score}: the weighted score from Equation~\eqref{eq:similarity}.
    \item \textbf{description}: the free-text visual description. % of the stored CAPTCHA.
    \item \textbf{tools\_applied}: the ordered list of preprocessing tools used in that run.
    \item \textbf{slot}: \texttt{correct} or \texttt{failed}, indicating the outcome.
    \item \textbf{note}: a natural-language annotation ``\checkmark~succeeded replicate if 
    features match'' or ``$\times$~failed avoid the same tools on similar images''.
    \item \textbf{ocr\_result}: the raw OCR output, included only for \texttt{failed} records 
    to assist diagnosis.
\end{itemize}

\noindent\textbf{Relevance thresholds.}
Table~\ref{tab:rag_policy} summarizes the threshold policy applied to retrieved RAG records. 
Correct-slot records are adopted, verified, or ignored according to their similarity to the 
query; failed-slot records serve as negative signals regardless of similarity, and a tool 
that appears repeatedly in failed records is treated as banned for that CAPTCHA style.

\begin{table}[h]
\centering
\caption{Planner policy for retrieved RAG records.}
\label{tab:rag_policy}
\scriptsize
\begin{tabular}{llp{4.5cm}}
\toprule
\textbf{Slot} & \textbf{Similarity} & \textbf{Action} \\
\midrule
\multirow{3}{*}{Correct}
  & $\geq 0.8$       & Adopt pipeline if features match \\
  & $0.6$--$0.8$     & Verify each feature before copying \\
  & $< 0.6$          & Ignore entirely \\
\midrule
\multirow{3}{*}{Failed}
  & $\geq 0.6$       & Avoid same tools \\
  & $0.5$--$0.6$     & Still a negative signal \\
  & Repeated         & Treat that tool as banned \\
\bottomrule
\end{tabular}
\end{table}

\noindent\textbf{Positive and negative records.}
A distinguishing feature of the RAG store is that it retains \emph{both} successful and failed attempts.
Failed records are explicitly valuable: when the Planner sees that a particular pipeline failed on a visually similar CAPTCHA, it avoids repeating the same strategy.
In the prompt, failed records are formatted as ``\textit{the following pipeline was tried on a similar image and FAILED: [...]}'' to clearly signal that the strategy should be avoided.

\subsection{Executor and Toolbox}
\label{subsec:executor}
 
The Executor receives the Planner's pipeline specification and runs it in three phases.

\noindent\textbf{Phase 1: Image Preprocessing.}
The toolbox provides five preprocessing operations to neutralise a specific class of visual degradation commonly encountered in CAPTCHA images.
Table~\ref{tab:tools} summarises the tools, their target degradation, and the
conditions under which each should be applied.

\begin{table}[h]
\centering
\caption{Preprocessing tools and their intended use cases.}
\label{tab:tools}
\small
\renewcommand{\arraystretch}{1.2}
\begin{tabular}{lp{5.5cm}}
\toprule
\textbf{Tool} & \textbf{Description and Usage Condition} \\
\midrule
\texttt{color\_extract} & Removes colored interference lines by isolating non-color pixels. Use when wavy or diagonal lines of a distinct color cross the text region. \\
\texttt{sharpen} & Restores character edge crispness via unsharp masking. Use when blur distortion has softened character boundaries. \\
\texttt{denoise} & Removes small connected components below an area threshold. Use to clean residual noise dots or on patterned backgrounds. \\
\texttt{remove\_line} & Detects and removes a single thin dark diagonal or near-horizontal line via the Hough transform. Use when such a line crosses all characters. \\
\texttt{binarize} & Converts a faded-grayscale image to black-and-white using Otsu's global threshold. Use \textit{only} when \texttt{background = "clean"}. \\
\bottomrule
\end{tabular}
\end{table}

\noindent Three hard ordering constraints govern pipeline composition and are surfaced as explicit rules in the Planner prompt:

\begin{enumerate}
    \item \texttt{color\_extract} must precede \texttt{sharpen} and
          \texttt{binarize}, as both destroy the color information on which
          color-based line removal relies.
    \item \texttt{denoise} is applied after \texttt{color\_extract}
          to eliminate residual dot artefacts produced by color separation.
    \item \texttt{binarization} is used only when \texttt{background = ``clean''};
          applying Otsu's global threshold to a structured or gradient background
          corrupts the result.
\end{enumerate}

\noindent The Planner may select zero, one, or several tools depending on the
visual analysis, and the tools are designed to be freely composable within these
constraints.
\begin{figure}[h]
\centering
\begin{subfigure}{0.42\columnwidth}
    \includegraphics[width=\linewidth]{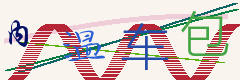}
    \caption*{Before}
\end{subfigure}
\hspace{0.05\columnwidth}
\begin{subfigure}{0.42\columnwidth}
    \includegraphics[width=\linewidth]{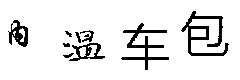}
    \caption*{After}
\end{subfigure}

\small(a) Color Extract on $\mathcal{D}_1$ Chinese CAPTCHA.
\vspace{0.2cm}

\begin{subfigure}{0.4\columnwidth}
    \includegraphics[width=\linewidth]{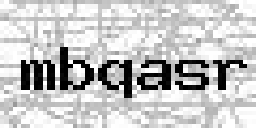}
    \caption*{Before}
\end{subfigure}
\hspace{0.05\columnwidth}
\begin{subfigure}{0.4\columnwidth}
    \includegraphics[width=\linewidth]{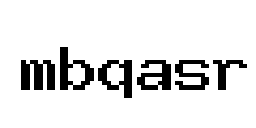}
    \caption*{After}
\end{subfigure}

\small(b) Denoise on $\mathcal{D}_4$
\caption{Effective use of preprocessing tools.}
\label{fig:preprocessing}
\end{figure} 

\begin{figure}[h]
\centering
\begin{subfigure}{0.42\columnwidth}
    \includegraphics[width=\linewidth]{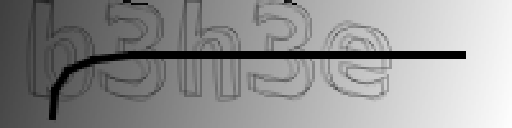}
    \caption*{Before}
\end{subfigure}
\hspace{0.05\columnwidth}
\begin{subfigure}{0.42\columnwidth}
    \includegraphics[width=\linewidth]{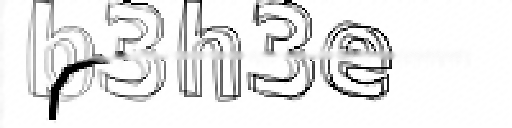}
    \caption*{After}
\end{subfigure}

\small(a) Remove Line on $\mathcal{D}_2$  

\vspace{0.2cm}
\begin{subfigure}{0.42\columnwidth}
    \includegraphics[width=\linewidth]{dataset/D2_before_remove_line.png}
    \caption*{Before}
\end{subfigure}
\hspace{0.05\columnwidth}
\begin{subfigure}{0.42\columnwidth}
    \includegraphics[width=\linewidth]{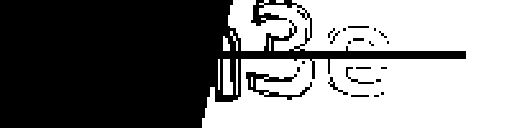}
    \caption*{After}
\end{subfigure}

\small(b) \textit{Failed} Binarize on $\mathcal{D}_2$ 

\caption{Effective vs.\ degenerative use of preprocessing tools.}
\label{fig:wrong_use}
\end{figure}

As shown in Fig.~\ref{fig:preprocessing}, for instance, \texttt{color\_extract} $\to$ \texttt{denoise} $\to$ \texttt{binarize} is a valid three-step pipeline for a clean-background CAPTCHA overlaid with colored lines and residual dot noise. 
Conversely, misapplied preprocessing operations cause more visual obfuscation, making the CAPTCHA even harder for the underlying VLM (see the comparison in Fig.~\ref{fig:wrong_use}).
These cases concretely illustrate design principle 4 (Sec.~\ref{sec:design}): the same tool can help one CAPTCHA archetype and harm another, so tool selection must be guided by retrieved success and failure records rather than applied uniformly. This is the trade-off the dual-signal RAG store (Sec.~\ref{subsec:rag}) is built to manage, and which we quantify in Sec.~\ref{sec:degen-recov}.

\noindent\textbf{Phase 2: OCR engine.}
After preprocessing, the Executor invokes the same VLM used for planning, now with a type-specific OCR prompt:
\begin{itemize}
    \item \textit{Alphanumeric prompt}: instructs the model to output exactly the characters visible in the image, handling mixed letters and digits.
    \item \textit{Math prompt}: instructs the model to extract the arithmetic expression as written (e.g., \texttt{3+5=?}), without solving it.
    \item \textit{Chinese prompt}: instructs the model to output each character independently, without inferring semantic meaning between characters.
\end{itemize}
The prompt type is determined by the \texttt{captcha\_type} field from the Planner's Stage~1 analysis.

\noindent\textbf{Phase 3: Post-processing.}
For math CAPTCHAs, the OCR output (e.g., \texttt{3+5=?}) is passed to a rule-based arithmetic solver.
Before evaluation, the system validates the expression: it must contain at least one arithmetic operator and an equals sign or question mark, and must not contain alphabetic characters.
If validation fails, the system reclassifies the CAPTCHA as alphanumeric and re-runs OCR with the alphanumeric prompt.
This fallback catches misclassifications where the Planner incorrectly identified a CAPTCHA as a math expression.

One additional recovery mechanism: If the Planner labels a CAPTCHA as \texttt{alphanumeric} but the OCR output contains \texttt{=} or \texttt{?}, the executor automatically attempts to solve the math before returning a result.

\subsection{Ground-truth Feedback Loop and Self-Improvement}
\label{subsec:feedback}
 
After each solve attempt, the Executor writes a complete record to the RAG Knowledge Base:
 
\begin{center}
\small
\texttt{\{image\_hash, description, pipeline, params, ocr\_output, outcome, timestamp\}}
\end{center}
 
\noindent This write occurs for both passed and failed attempts.
The resulting feedback loop drives \sol's self-improvement through three mechanisms:

\noindent\textbf{Positive reinforcement.}
When a pipeline succeeds, future CAPTCHAs with similar visual properties will retrieve this record as a high-confidence reference, biasing the Planner toward the same strategy.
Over consecutive images from the same site, successful strategies are reinforced and converge.
 
\noindent\textbf{Negative avoidance.}
When a pipeline fails, the record is stored with a \texttt{FAIL} label.
Future retrievals surface this record alongside successful ones, explicitly informing the Planner that the failed strategy should be avoided.
This is particularly important for addressing the degeneration-recovery trade-off: if a preprocessing step degraded a similar image in the past, the Planner learns to skip it.
 
\noindent\textbf{Cold start and warm-up.}
When \sol first encounters a new marketplace, the RAG store contains no records for that site's CAPTCHA style.
In this cold-start scenario, the Planner relies solely on its own visual analysis (Stage~1) without retrieved experience.
As attempts accumulate, typically within 5-10 images as per our observation, the RAG store builds sufficient site-specific history for retrieval to become effective.

\textbf{Cross-site adaptation.}
Because retrieval is based on visual similarity rather than site identity, experience can be shared across marketplaces with similar CAPTCHA characteristics. We empirically evaluate this capability
using the mixed-site experiment in Sec.~\ref{sec:cross-site}.

\section{Experiment Setup}

\subsection{Models and Hardware Configurations}
\label{sec:models}
We evaluate nine models across three categories: one OCR specialist, five general-purpose instruct models, and three general-purpose thinking models.
% We evaluate eight models across three categories: one OCR specialist, five general-purpose instruct models, and two general-purpose thinking models.

HunyuanOCR~\cite{hunyuanocr} is a 1B end-to-end OCR specialist covering multiple languages and document types. 
It is the lowest-cost point in our comparison and represents the OCR-specialist category.

The Qwen3-VL family~\cite{qwen3vl2025} contributes three instruct models at 4B, 8B, and 30B-A3B. The first two are dense; the third model is a mixture-of-experts model with 3B active parameters per forward pass.

Gemma4-E4B-it and Gemma4-12B-it~\cite{gemma4} are instruct models having 8B total parameters with 4.5B effective and 12B parameters (all effective), respectively.

Qwen3.5-9B~\cite{qwen35} and Qwen3.6-27B~\cite{qwen36} are native multimodal thinking models that integrate vision and language through early fusion.

Step3-VL-10B~\cite{step3vl2026} is a 10B thinking model pairing a Perception Encoder with a Qwen3-8B backbone. 
It introduces Parallel Coordinated Reasoning (PaCoRe) for test-time compute scaling.

All experiments were conducted on a single NVIDIA RTX Pro 6000 GPU (96 GB VRAM) running Debian 13 (trixie) with NVIDIA driver 610.57.04. 
To simulate a resource-constrained production environment, we capped GPU memory at 30\% of its total capacity (~29 GB), using vLLM (v0.25.1) to enforce this limit. 
All models were evaluated in their official, unmodified form: Qwen3-VL and Qwen3.5 (Qwen), Gemma4 (Google), and Step3-VL (StepFun). 
No third-party quantization, fine-tuning, or other modifications were applied. 

\subsection{Datasets}
\label{sec:datasets}
 
We evaluate \sol on four CAPTCHA datasets collected from real dark web marketplaces, namely ChangAn ($\mathcal{D}_1$), Freecity ($\mathcal{D}_2$), Russian market ($\mathcal{D}_3$), Mobilestore ($\mathcal{D}_4$).
CAPTCHA images were collected via automated visits to each marketplace's login page, rate-paced with randomized delays to stay below detection thresholds. Each visit yielded a fresh CAPTCHA, which was downloaded and stored.
% Table~\ref{tab:datasets} provides a summary.
For each marketplace, we sampled 500 CAPTCHA instances, resulting in 2,000 images in total.
Table~\ref{tab:datasets} summarizes the resulting dataset structure and visual characteristics.
% All images were collected through Tor browser sessions during routine CTI monitoring.
Dataset availability and responsible-use safeguards are discussed in Appendix~\ref{sec:opensource}.

\begin{table*}[t]
\centering
\caption{Summary of dataset visual properties.}
\label{tab:datasets}
\resizebox{1.8\columnwidth}{!}{%
\renewcommand{\arraystretch}{1.2}
\begin{tabular}{lllllllll}
\toprule
\textbf{Dataset} & \textbf{Type} & \textbf{Sample size} & \textbf{Background} 
& \textbf{Noise} & \textbf{Distortion} 
& \textbf{Char Quality} & \textbf{Difficulty}  & \textbf{Dark web site}\\
\midrule
$\mathcal{D}_1$         & Alpha / Math / Chinese & 500 & Mixed  
& colored lines      & Wave & Mixed   & High & ChangAn\\
$\mathcal{D}_2$        & Alphanumeric   & 500     & Clean     
& Strikethrough line  & None & Normal  & Low & Freecity\\
$\mathcal{D}_3$  & Alphanumeric   & 500     & Gradient  
& Dots/specks         & Wave & Normal  & Medium & Russian market\\
$\mathcal{D}_4$    & Alphanumeric & 500       & Patterned 
& None (mesh texture) & None & Thick   & Medium & Mobilestore \\

\bottomrule
\end{tabular}%
}
\end{table*}

\begin{figure}[h]
\centering
% Row 1
\begin{minipage}[t]{0.46\columnwidth}
  \centering
  \includegraphics[width=\linewidth]{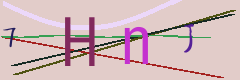}
  \small\textbf{(a) $\mathcal{D}_1$}\\
  \scriptsize Alphanumeric, colored\\
  interference lines, wave distortion.
\end{minipage}
\hfill
\begin{minipage}[t]{0.48\columnwidth}
  \centering
  \includegraphics[width=\linewidth]{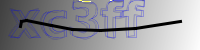}
  \small\textbf{(b) $\mathcal{D}_2$}\\
  \scriptsize Alphanumeric, clean background,\\
  single strikethrough line.
\end{minipage}

% Row 2
\begin{minipage}[t]{0.45\columnwidth}
  \centering
  \includegraphics[width=\linewidth]{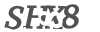}
  \small\textbf{(c) $\mathcal{D}_3$}\\
  \scriptsize Alphanumeric, gradient background,\\
  wave distortion and dot noise.
\end{minipage}
\hspace{0.05\columnwidth}
\begin{minipage}[t]{0.4\columnwidth}
  \centering
  \includegraphics[width=\linewidth]{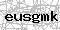}
  \small\textbf{(d) $\mathcal{D}_4$}\\

  \scriptsize Alphanumeric, patterned mesh\\
  background, 60$\times$30 px.
\end{minipage}

\caption{Representative samples from each dataset.}
\label{fig:samples}
\end{figure}

\noindent\textbf{$\mathcal{D}_1$} is the most diverse dataset, containing CAPTCHAs spanning three types: alphanumeric, math equation, and Chinese characters.
The images exhibit a range of visual defenses, e.g., colored interference lines, wave-distorted characters, overlapping strokes, and mixed background textures, making it the primary benchmark for evaluating the Planner's ability to select appropriate tools for heterogeneous inputs.

\noindent\textbf{$\mathcal{D}_2$} contains alphanumeric CAPTCHAs with clean backgrounds and a dark line crossing the text.
This dataset specifically tests whether \sol correctly applies minimal or \emph{no} preprocessing to easy images.
A well-functioning Planner should recognize that preprocessing is unnecessary here and leave the image untouched, directly evaluating the system's handling of the degeneration-recovery effect. 
% \todo{we are mentioning Gap 2 here; can we refer to other gaps at appropriate places in evaluations?}

\noindent\textbf{$\mathcal{D}_3$} contains alphanumeric CAPTCHAs with moderate noise and distortion, serving as an intermediate-difficulty benchmark between $\mathcal{D}_2$ and $\mathcal{D}_1$.
 
\noindent\textbf{$\mathcal{D}_4$} contains alphanumeric CAPTCHAs with thick characters on a patterned mesh background. 

\noindent\textbf{Dataset Annotation and Verification.}
\label{sec:verification}
\noindent Ground-truth labels were produced independently prior to the model experiments, ensuring that annotation was not influenced by model predictions. 
Three annotators participated in the annotation process %, two of whom are Chinese-speaking and independently verified the Chinese-script CAPTCHAs. 
and inter-annotator reliability was measured before any reconciliation using pairwise exact-string agreement (ESA) and character error rate (CER). 
% ESA measures whether two independently produced transcriptions match exactly, while CER quantifies character-level disagreement using normalized edit distance.
Disagreements were subsequently reviewed and manually resolved through consensus to establish the final ground-truth labels.
More details and dataset-level agreement statistics are reported in Appendix~\ref{app:annotation}. %, Table~\ref{tab:annotation_agreement}.}

% Ground-truth labels were obtained by manual transcription of the visible CAPTCHA characters. Labels were checked by manual review rather than a formal inter-annotator study; for the Chinese CAPTCHAs in D1, three Chinese-speaking reviewers checked the transcriptions and resolved disagreements by discussion. We did not compute formal agreement statistics, so some residual labeling noise may remain.

\subsection{Baselines and Models}
\label{sec:baselines}

We evaluate nine open-weight models, with model size spanning from 1B to 30B parameters (Table~\ref{tab:baseline_simple}).
Each model is tested in two configurations:

\noindent\textbf{Baseline (raw VLM).}
A CAPTCHA image is passed directly to the model with one generic prompt for all types.
There is no preprocessing, no type-specific prompting, and no RAG retrieval involved.
This set of evaluations demonstrates what each model achieves on its own.

\noindent\textbf{\sol (agentic pipeline).}
The \sol system is applied on top of the same model, with visual analysis, RAG retrieval, selective preprocessing, type-specific prompting, and feedback recording.
%The difference between the two configurations isolates the accuracy gain attributable entirely to agentic tool selection without any additional training.
%HunyuanOCR is the sole exception---as a pure OCR system without visual-reasoning capability, it cannot serve as a Planner and is evaluated in baseline only.

Throughout this paper, \emph{backbone} refers to the underlying base VLM on top of which the \sol agentic pipeline is applied.

\begin{table}[htbp]
\centering
\caption{Baseline models evaluation: Parameter size, type, domain, and development source.}
\label{tab:baseline_simple}
\resizebox{\columnwidth}{!}{%
\begin{tabular}{l c l c c c l}
\toprule
\textbf{Model Name} & \textbf{Size} & \textbf{Mode}  & \textbf{Domain} & \textbf{Date} & \textbf{Source} \\
\midrule
HunyuanOCR & 1B &\texttt{Inst}  & OCR & Nov 2025 & Tencent \\
Gemma4-E4B-it & 8B  & \texttt{Inst} &General & Apr 2026 & Google \\
Qwen3-vl-4B  & 4B & \texttt{Inst}  &General & Oct 2025 & Alibaba \\
Qwen3-vl-8B  & 8B & \texttt{Inst} &General & Oct 2025 & Alibaba \\
Qwen3.5-9B   & 9B  & \texttt{Think} &General & Mar 2026 & Alibaba \\
Step3-VL-10B & 10B & \texttt{Think} &General & Jan 2026 & Stepfun \\
Gemma4-12B-it & 12B  & \texttt{Inst} &General & Jul 2026 & Google \\
Qwen3.6-27B & 27B &  \texttt{Think} &General & Apr 2026 & Alibaba \\
Qwen3-VL-30B-A3B  & 30B & \texttt{Inst} &General & Oct 2025 & Alibaba \\
\bottomrule
\end{tabular}%
}
\end{table}

Eight of the nine evaluated models are general-purpose VLMs. 
We include HunyuanOCR as our sole OCR-specialized baseline to determine whether dedicated text-recognition fine-tuning is sufficient on its own. 
The selected models span 1B to 30B parameters across four distinct architectures, to understand if our findings generalize across different model families and training paradigms.
A key distinction among these models is their \emph{inference mode} (Table~\ref{tab:baseline_simple}). 
Instruct-only models (marked \texttt{Inst}) execute structured commands without internal chain-of-thought deliberation, prioritizing speed over planning reliability. 
Conversely, thinking-enabled models (marked \texttt{Think}) perform extended internal reasoning, yielding more deliberate tool-selection decisions at the expense of higher latency. 
This dichotomy is critical: \sol requires the backbone to both \emph{plan} (reason about tool selection) and \emph{read} (execute OCR). 
These are two distinct tasks that inherently benefit from different inference modes.

\subsection{Evaluation Metrics}
\label{sec:metrics}
 
\noindent\textbf{Exact-match accuracy.}
The primary metric: the fraction of CAPTCHAs for which the website returns ``pass'' upon submission.
Matching is case-insensitive for alphabetic characters, following standard practice~\cite{captcha-Usenix-2025}, which we also experimented on the markets we gathered our dataset from.
For math CAPTCHAs, the numeric result must match exactly.

\noindent\textbf{Inference time.}
Wall-clock time per CAPTCHA on local hardware, covering all pipeline stages (visual analysis, RAG query, preprocessing, OCR, and result handling). This measures local compute only; the Tor network round-trip is excluded.

\noindent\textbf{Degeneration and recovery rates}.
For each preprocessing tool, we measure:
(i)~\textit{degeneration rate}: fraction of images correct without the tool but incorrect with it (correct $\to$ wrong);
(ii)~\textit{recovery rate}: fraction incorrect without the tool but correct with it (wrong $\to$ correct).
Statistical significance is assessed via McNemar's test ($p < 0.05$).

\section{Results and Analyses}
\label{sec:results}

\subsection{CAPTCHA-solving Accuracy}
\begin{figure}
\centering
    \vspace{-1em}
    \includegraphics[width=0.5\textwidth]{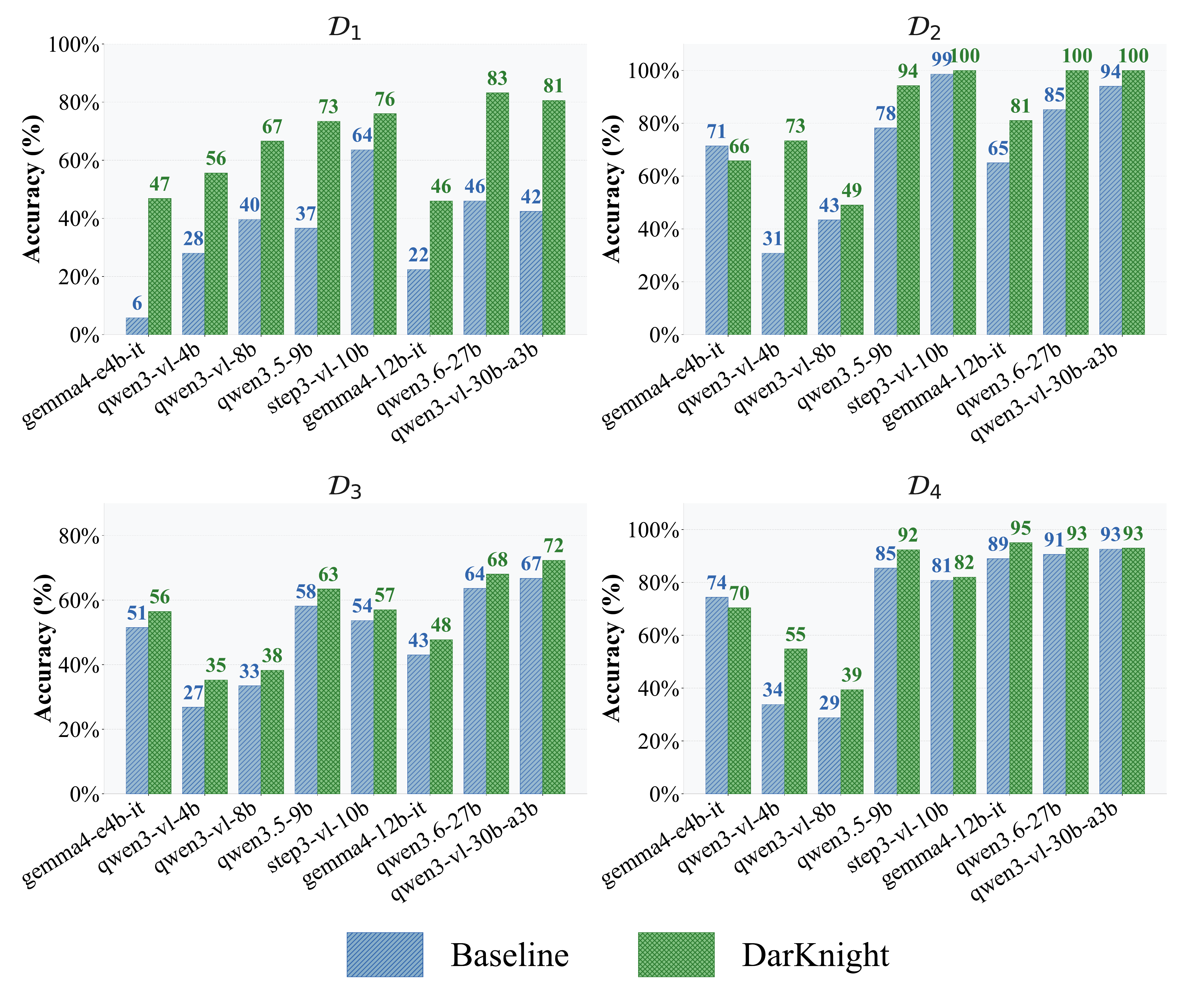}
\caption{Baseline vs. \sol: accuracy (\%) across four benchmark datasets.}
\label{fig:accuracy_charts}
\end{figure}

Fig.~\ref{fig:accuracy_charts} presents the full empirical results across all four datasets; the corresponding exact numerical values are compiled in Table~\ref{tab:main-results} (Appendix~\ref{sec:accuracytab}).

\noindent\textbf{Baseline models performance.}
Without any agentic assistance, open-weight VLMs exhibit highly variable performance across dark web CAPTCHA datasets.
On $\mathcal{D}_1$, the most diverse dataset, containing alphanumeric, math, and Chinese CAPTCHAs with heavy visual defenses, baseline accuracy is strikingly low, ranging from 6\% with Gemma4-e4b-it (see Table~\ref{tab:main-results}) to 64\% (achieved by Step3-VL-10B).
Most general-purpose VLMs cluster around 22--46\%, indicating that dark web CAPTCHAs with multiple defense features remain challenging for current models. 
Conversely, on the less defended alphanumeric CAPTCHAs of $\mathcal{D}_4$, several models exceed 80\% baseline accuracy. 
Models also perform reliably on $\mathcal{D}_2$'s clean, minimal-noise backgrounds (Step3-VL-10B: 98.6\%). 
$\mathcal{D}_3$ represents an intermediate difficulty, with baselines falling between 27--67\%.

\noindent\textbf{Model scale does not predict accuracy}. 
On $\mathcal{D}_1$, parameter count is a poor predictor of baseline accuracy. The dense Qwen3.6-27B (46.0\%) improves on Qwen3-VL-8B (39.6\%) by only 6.4~pp despite a 3.4$\times$ larger parameter count, and despite also being a newer model with built-in reasoning. The mixture-of-experts Qwen3-VL-30B-A3B (42.4\%), which activates only 3B parameters per token, performs on par with its dense 8B sibling.
Conversely, Step3-VL-10B reaches 63.6\% on $\mathcal{D}_1$, outperforming every larger open-weight model we evaluate, while Gemma4-12B-IT (22.4\%) falls below the 4B Qwen3-VL (28.0\%).
However, pure OCR systems lacking visual reasoning struggle on complex datasets. HunyuanOCR (cf.~Table~\ref{tab:main-results}) achieves 90.0\% on the simple $\mathcal{D}_4$ but collapses to 12.4\% on $\mathcal{D}_1$.

\noindent\textbf{Agentic Tool Selection vs.\ Model Scaling.}
Beyond baseline performance, in our evaluation, \sol's agentic framework yields higher accuracy gains than scaling model parameters.
Within the dense Qwen3-VL family, doubling parameters from 4B to 8B raises $\mathcal{D}_1$ accuracy by 11.6~pp
(28.0\%$\rightarrow$39.6\%), whereas adding \sol to the 4B model raises it by 27.6~pp (to 55.6\%), so the 4B model with \sol exceeds the base 8B by 16~pp.
% On $\mathcal{D}_1$, scaling the Qwen3-VL backbone from 8B to 30B improves accuracy only by +2.8\% (39.6\%$\to$42.4\%). 
% In contrast, integrating \sol with the 8B model boosts its accuracy by a significant +27\%, an improvement ~10$\times$ larger than scaling. 
This improvement holds across model families: even Qwen3-VL-30B-A3B rises from 42.4\% to 80.6\% with \sol (+38.2~pp). 
Gemma4-E4B-IT with \sol{} (46.8\%) also slightly exceeds the bare Qwen3-VL-30B-A3B (42.4\%).

We additionally evaluate two conventional OCR systems and four frontier cloud VLMs as reference baselines; although the latter remains outside our target deployment setting as outlined in Sec.~\ref{sec:intro}. 
In the first group, PaddleOCR~\cite{paddleocr2020} substantially outperforms Tesseract~\cite{smith2007tesseract}, particularly on the less distorted datasets, but remains considerably weaker on the heavily defended $\mathcal{D}_1$ dataset (below 5\%). 
Recent frontier API models from Claude should provide stronger general-purpose visual reasoning, although their performance was below expectations; even with the Claude Opus 5, the accuracy is less than 50\% (see Appendix~\ref{app:additional_baselines} for more details).

\subsection{The Degeneration-Recovery Effect}
\label{sec:degen-recov}

A central design premise of \sol is that preprocessing must be applied selectively rather than uniformly. 
To validate this, we trace every tool-selection decision made by the Planner across the evaluation suite and classify its ultimate effect on the prediction. 
By comparing the \sol pipeline's outcome against the raw VLM baseline, we map each attempt to one of the four transition states: correct$\to$correct (C$\to$C, preserving a correct prediction), correct$\to$wrong (C$\to$W, \emph{degeneration}), wrong$\to$correct (W$\to$C, \emph{recovery}), and wrong$\to$wrong (W$\to$W, no effect). 
Based on these transition states, we derive three core metrics to quantify the Planner's efficacy: \textbf{Degen\%} (the fraction of originally correct baseline predictions degraded by unnecessary tools), \textbf{Recov\%} (the fraction of originally incorrect baseline predictions successfully recovered), and \textbf{Net} (the overall performance gain, calculated as Recov\% $-$ Degen\%).

\begin{table}[htbp]
\centering
\caption{Per-tool degeneration-recovery profile across all models and datasets. Tools ordered by net effect.}
\label{tab:tool-risk}
\resizebox{1.0\columnwidth}{!}{%
\renewcommand{\arraystretch}{1.15}
\begin{tabular}{l r r r r r r r r r}
\toprule
\textbf{Tool}
  & \textbf{N} & \textbf{\%}
  & \textbf{C$\to$C} & \textbf{C$\to$W}
  & \textbf{W$\to$C} & \textbf{W$\to$W}
  & \textbf{Degen\%} & \textbf{Recov\%}
  & \textbf{Net} \\
\midrule
(no tool)
  & 4406 & 72.8 & 2357 & 159 & 733  & 1157
  & 6.3  & 38.8 & $+$32.5 \\
color\_extract
  & 1216 & 20.1 & 371  & 58  & 508  & 279
  & 13.5 & 64.5 & $+$51.0 \\
sharpen
  & 197  & 3.3  & 66   & 3   & 90   & 38
  & 4.3  & 70.3 & $+$66.0 \\
remove\_straight\_line
  & 184  & 3.0  & 111  & 9   & 53   & 11
  & 7.5  & 82.8 & $+$75.3 \\
denoise
  & 43   & 0.7  & 17   & 12  & 6    & 8
  & 41.4 & 42.9 & $+$1.5  \\
binarize
  & 3    & 0.0  & 0    & 0   & 0    & 3
  & ---  & 0.0  & ---     \\
\midrule
\rowcolor{gray!15}
\textbf{Total}
  & \textbf{6049} & \textbf{100}
  & \textbf{2922} & \textbf{241}
  & \textbf{1390} & \textbf{1496}
  & \textbf{7.6} & \textbf{48.2}
  & \textbf{$+$40.5} \\
\bottomrule
\end{tabular}%
}
\end{table}

Table~\ref{tab:tool-risk} summarises the effect of each preprocessing tool applied by \sol across three operational models (Qwen3-VL-4B, Qwen3.5-9B, and Qwen3-VL-30B-A3B; see Sec.~\ref{sec:cross-model}) and all datasets. 
The bottom row reports the pooled degeneration and recovery rates across all Planner decisions, including cases in which no preprocessing tool was selected.
% The degeneration and recovery rates in the bottom row of the table report the count-weighted average, where the weights are directly proportional to the invocation frequency of each individual tool in the pipeline.

Across all Planner decisions, \sol recovers 48.2\% of baseline failures at a cost of 7.6\% degeneration, yielding 1,390 recoveries against only 241 degenerations (a \textbf{5.8$\times$} recovery-to-degeneration ratio).
% When the Planner does apply a tool, it recovers 70.7\% of baseline failures at a cost of 7.8\% degeneration,
% yielding 1{,}340 recoveries against only 216 degenerations (a \textbf{6.2$\times$} recovery-to-degeneration ratio).

The preprocessing tools exhibit different risk-reward profiles.
\texttt{remove\_straight\_line} achieves the strongest net effect ($+$75.3), recovering 82.8\% of baseline failures while degrading 7.5\% of correct predictions. 
\texttt{sharpen} provides the next strongest profile ($+$66.0), with a low degeneration rate of 4.3\%
and a recovery rate of 70.3\%. \texttt{color\_extract}, the most frequently selected preprocessing tool, achieves a $+$51.0 net effect, recovering 64.5\% of failures at a 13.5\% degeneration rate. 
In contrast, \texttt{denoise} provides only a marginal net benefit ($+$1.5), while \texttt{binarize} is selected only three times and therefore does not provide sufficient evidence for a meaningful risk-reward estimate.

\begin{table}[htbp]
\centering
\caption{Planner tool-selection behavior per dataset.
\textbf{No-preproc rate} is the fraction of decisions where the Planner selects an empty pipeline.}
\label{tab:no-preproc-rate}
\resizebox{\columnwidth}{!}{%
\renewcommand{\arraystretch}{1.15}
\begin{tabular}{l c c c c}
\toprule
& \textbf{$\mathcal{D}_1$} & \textbf{$\mathcal{D}_2$}
& \textbf{$\mathcal{D}_3$} & \textbf{$\mathcal{D}_4$} \\
\midrule
No-preproc rate
& 26.6\% & 86.9\% & 97.0\% & 56.7\% \\
Avg baseline acc
& 35.7\% & 67.7\% & 50.5\% & 70.6\% \\
Avg $\Delta$ with \sol
& +34.2 & +21.5 & +6.5 & +9.5 \\
\bottomrule
\end{tabular}%
}
\end{table}

Table~\ref{tab:no-preproc-rate} demonstrates that the Planner's intervention behavior varies substantially across datasets.
On $\mathcal{D}_3$, the Planner selects no preprocessing in 97.0\% of cases, while the no-preprocessing rate remains high at 86.9\% on $\mathcal{D}_2$. Conversely, on $\mathcal{D}_1$, which contains substantially heavier visual obfuscation, the Planner intervenes in 73.4\% of cases. 
This indicates that tool selection is driven by the visual characteristics of the CAPTCHA rather than by a uniform preprocessing policy.

% Table~\ref{tab:no-preproc-rate} demonstrates that the Planner's intervention behavior varies substantially across datasets. 
% On $\mathcal{D}_3$, the Planner selects no preprocessing in 97.0\% of cases, while the no-preprocessing rate remains high at 86.9\% on $\mathcal{D}_2$. 
% Conversely, on $\mathcal{D}_1$, which contains substantially heavier visual obfuscation, the Planner intervenes in 73.4\% of cases. This indicates that tool selection is driven by the visual characteristics of the CAPTCHA rather than by a uniform preprocessing policy.

% Across the entire evaluation, \sol produces 1{,}390 recoveries against 241 degenerations, yielding a \textbf{5.8$\times$} recovery-to-degeneration ratio.
Importantly, selecting no preprocessing is itself an explicit Planner decision.
When intervention is unnecessary, withholding tools avoids potential degeneration. 
Overall, in our evaluation,  effective planning required both selecting appropriate transformations and knowing when not to intervene.

\subsection{Effect of Retrieval Memory}
\label{sec:rag_ablation}

\begin{figure}[htbp]
    \centering
    \vspace{-1em}
    \includegraphics[width=0.5\textwidth]{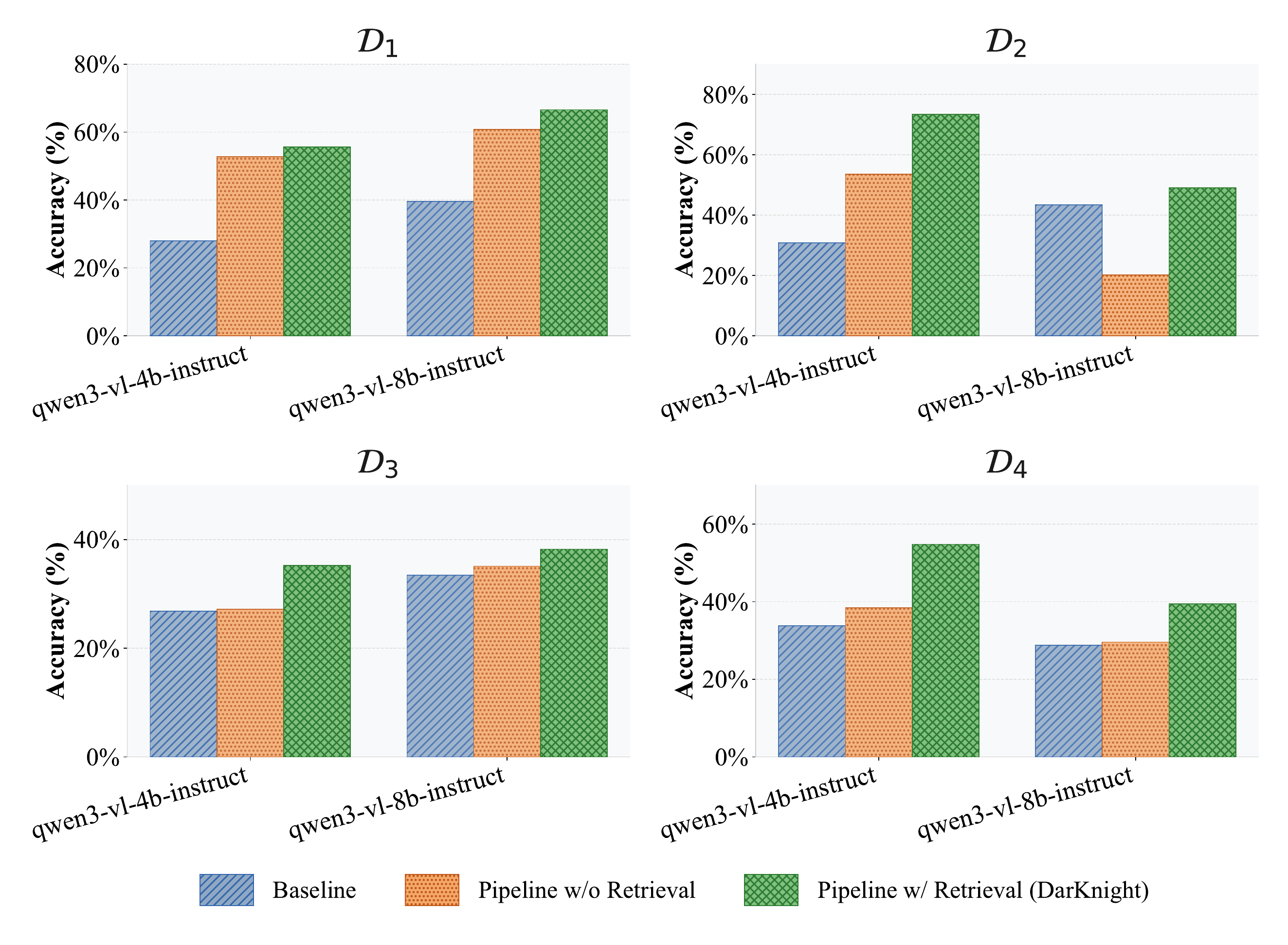}
    \caption{Ablation of retrieval memory using Qwen3-VL-4B and Qwen3-VL-8B across $\mathcal D_1 - \mathcal D_4$. 
    % We compare the raw backbone, the agentic pipeline without retrieval, and the full \sol pipeline with retrieval.
    }
    \label{fig:rag_ablation}
\end{figure}

To isolate the contribution of retrieval memory, we compare the raw VLM baseline, the \sol pipeline without retrieval, and the full \sol framework with retrieval. 
As shown in Fig.~\ref{fig:rag_ablation}, enabling retrieval consistently improves
accuracy over the no-retrieval pipeline across all evaluated configurations. 
This demonstrates that the gains of \sol are not solely attributable to preprocessing and prompting.
Retrieving prior successful and failed experiences provides additional benefit by guiding tool selection and avoiding previously harmful actions. 
We further examine the sensitivity to RAG memory size in App.~\ref{app:rag_size}, where a memory size of 100 achieves the best average performance across the evaluated settings and is adopted as the default configuration.

\subsection{Cross-Site Adaptation}\label{sec:cross-site}

To evaluate whether \sol can operate across heterogeneous marketplaces using a shared experience memory, we construct a mixed-site stream from all four datasets.
Specifically, we sample 125 CAPTCHAs from each of $\mathcal{D}_1$ to $\mathcal{D}_4$, combine them into a merged set with 500 images, and randomly shuffle the samples. 
The resulting stream is processed sequentially using a single shared RAG store, allowing experience accumulated from one marketplace to be retrieved when solving visually similar CAPTCHAs originating from another.
Table~\ref{tab:cross_site} reports the results. \sol improves accuracy for all four evaluated backbones in this mixed-site setting. Qwen3-VL-8B improves from 65.0\% to 75.0\%, while
Qwen3.6-27B improves from 65.0\% to 78.0\%. 
We also observe substantial improvements for Gemma4-E4B and Qwen3.5-9B.
These results indicate that \sol can maintain a shared experience memory across heterogeneous CAPTCHA sources rather than requiring an independently engineered solver for each marketplace.

\begin{table}[htbp]
\centering
\caption{Accuracy (\%) in the cross-site mixed-stream experiment.}
\label{tab:cross_site}
\begin{tabular}{lccc}
\hline
\textbf{Model} & \textbf{Baseline} & \textbf{\sol} & $\Delta$ \\
\hline
Qwen3-VL-8B   & 65.0 & 75.0 & +10.0 \\
Gemma4-E4B    & 50.8 & 60.8 & +10.0 \\
Qwen3.5-9B    & 16.5 & 80.4 & +63.9 \\
Qwen3.6-27B   & 65.0 & 78.0 & +13.0 \\
\hline
\end{tabular}
\end{table}

\subsection{Inference Time and Computational Overhead}
\label{sec:inference-time}
 
Table~\ref{tab:inference-time-results} presents the end-to-end inference time across all datasets. 
% As expected, integrating the \sol agentic pipeline inherently increases latency compared to raw baselines. 
% This delay represents the computational cost of the perceive-plan-execute loop, which involves dual VLM invocations and intermediate RAG retrieval. 
As expected, integrating the \sol agentic pipeline inherently increases latency compared to raw baselines. 
This delay represents the computational cost of the perceive-plan-execute loop, which involves dual VLM invocations and intermediate RAG retrieval. 
However, this overhead varies substantially across the evaluated VLMs.

\begin{table}[htbp]
\centering
\caption{Mean end-to-end inference time (in seconds) per CAPTCHA across dark web datasets.}
\label{tab:inference-time-results}
\resizebox{\columnwidth}{!}{%
\begin{tabular}{l c | c c | c c | c c | c c}
\toprule
& & \multicolumn{2}{c|}{\textbf{$\mathcal{D}_1$}} & \multicolumn{2}{c|}{\textbf{$\mathcal{D}_4$}} & \multicolumn{2}{c|}{\textbf{$\mathcal{D}_2$}} & \multicolumn{2}{c}{\textbf{$\mathcal{D}_3$}} \\
\textbf{Model} & \textbf{Params} & \textbf{B} & \textbf{DK} & \textbf{B} & \textbf{DK} & \textbf{B} & \textbf{DK} & \textbf{B} & \textbf{DK} \\
\midrule
HunyuanOCR                & 1B  & 1.82  & --    & 0.10  & --    & 0.07  & --    & 0.13  & --    \\
\midrule
Gemma4-E4B-IT             & 8B  & 0.156 & 3.81  & 0.154 & 3.25  & 0.137 & 3.12  & 0.157 & 2.97  \\
Qwen3-VL-4B-Instruct      & 4B  & 0.122 & 2.72  & 0.102 & 2.21  & 0.120 & 2.57  & 0.108 & 2.34  \\
Qwen3-VL-8B-Instruct      & 8B  & 0.170 & 4.26  & 0.164 & 3.72  & 0.273 & 3.89  & 0.165 & 3.67  \\
Qwen3.5-9B                & 9B  & 0.168 & 5.38  & 0.162 & 4.70  & 0.163 & 5.17  & 0.166 & 4.43  \\
Step3-VL-10B              & 10B & 4.500 & 68.19 & 1.026 & 44.50 & 2.203 & 48.20 & 0.971 & 50.11 \\
Gemma4-12B-IT             & 12B & 0.156 & 4.44  & 0.164 & 5.47  & 0.144 & 4.03  & 0.172 & 3.77  \\
Qwen3.6-27B               & 27B & 2.247 & 23.03 & 1.714 & 18.34 & 1.898 & 21.00 & 1.732 & 20.70 \\
Qwen3-VL-30B-A3B-Instruct & 30B & 0.170 & 2.92  & 0.155 & 2.11  & 0.148 & 3.16  & 0.162 & 2.38  \\
\bottomrule
\end{tabular}%
}
\end{table}

\noindent\textbf{The Accuracy-Latency Sweet Spot.}
Among the evaluated models, Qwen3-VL-30B-A3B-Instruct provides the strongest accuracy--latency trade-off. 
With \sol, it achieves an average accuracy of 86.5\% across the four datasets while requiring only 2.64\,s per CAPTCHA
on average. 
In comparison, Qwen3.6-27B reaches a similar average accuracy of 86.1\% but requires 20.77\,s, while Step3-VL-10B
achieves 78.8\% at 52.75\,s. 
These results indicate that efficient model architecture and agentic orchestration matter more than model size alone.

% \noindent\textbf{The Lightweight Sweet Spot.} \levi{Binghui: the accuracy numbers are wrong! need to wait for Table X to be updated then find the sweet spot}
% For the instruct-tuned Qwen3-VL series (4B, 8B, 30B), the latency overhead is tightly bounded and highly viable for deployment. 
% \hlt{Among the lightweight instruct-tuned Qwen3-VL models, Qwen3-VL-8B provides the best accuracy–latency trade-off, requiring 3.60 s on $\mathcal{D}_1$ while reaching 85.4\% accuracy.
% We report this latency as the local computational overhead of the agentic pipeline; crawler request pacing and Tor network delay are separate operational factors.}
% Since marketplace session timeouts typically span 10 to 30 seconds, a sub-4-second inference time sits comfortably within this budget, avoids rate-limit triggers, and leaves margin for the Tor round-trip. 
% It guarantees a solve before session expiration while avoiding rate-limit triggers.

\noindent\textbf{The Latency Bottleneck of Heavy Reasoning.}
Conversely, some models incur higher local inference cost when integrated with \sol. 
Step3-VL-10B is the most expensive, requiring 44.50--68.19\,s per CAPTCHA across the four datasets,
with an average of 52.75\,s. Qwen3.6-27B also incurs considerably higher latency, averaging 20.77\,s, despite achieving strong accuracy. 
In contrast, Qwen3-VL-30B-A3B-Instruct achieves comparable or higher accuracy with an average latency of only 2.64\,s. 
These results show that the agentic overhead depends on model architecture and reasoning behavior rather than parameter count.

For the backbones we evaluated, effective CAPTCHA solving did not depend on parameter scaling alone. 
Instead, the best accuracy-latency trade-off came from combining an efficient VLM backbone with selective agentic orchestration.
 
% \noindent\textbf{The Latency Bottleneck of Heavy Reasoning.} \hlt{Conversely, pairing \sol with larger or thinking-enabled models introduces higher local inference overhead}. For instance, Qwen3.6-35B-A3B and Qwen3.5-9B demand 49.84s and 47.85s respectively on $\mathcal{D}_1$. This extreme degradation occurs because models optimized for extended internal reasoning generate excessive token sequences during the planning phase. Step3-VL-10B similarly struggles with a 25.34s latency. \hlt{Such delays make these models more computationally expensive for time-sensitive crawling workloads compared with the lightweight instruct models.}
 
% Ultimately, this latency analysis reinforces our core thesis: the optimal strategy for dark web CAPTCHA solving is not to scale up the parameter count or rely on heavy internal reasoning, but to orchestrate lightweight, instruction-following models (e.g., 8B) with an efficient agentic visual framework.

% \begin{obsbox}
% \textbf{Observations.}
% \begin{obsitem}
%     \item \hlt{Lightweight instruct models incur substantially lower agentic overhead than thinking-enabled models.}
%     %, ruling them out for unattended deployment.
%     \item \hlt{Latency is reported as a computational trade-off rather than a fixed deployment threshold.}
% \end{obsitem}
% \end{obsbox}

\subsection{Cross-Model Accuracy-Latency Analysis}
\label{sec:cross-model}

Table~\ref{tab:model-summary} shows that the benefit introduced by
\sol varies substantially across backbones and is not determined
simply by model size or inference mode. Qwen3-VL-4B obtains the
largest incremental improvement ($+$24.9\%), showing
that a relatively weak raw backbone can benefit substantially from
agentic assistance. In contrast, Qwen3-VL-30B-A3B-Instruct achieves
the highest final average accuracy (86.5\%), closely followed by
Qwen3.6-27B (86.1\%). These results also highlight the distinction
between pipeline-induced gain and absolute performance, a large
improvement does not necessarily imply the highest final accuracy.
\begin{table}[htbp]
\centering
\caption{Cross-model accuracy and latency summary.
Avg B and Avg DK denote the average raw-baseline and \sol
accuracy across the four datasets; Avg $\Delta$ denotes the
incremental improvement introduced by \sol, and $t_{\mathrm{DK}}$
denotes the average local inference time per CAPTCHA.}
\label{tab:model-summary}
\resizebox{\columnwidth}{!}{%
\renewcommand{\arraystretch}{1.15}
\begin{tabular}{l c r r r r}
\toprule
\textbf{Model} & \textbf{Params}
& \textbf{Avg B} & \textbf{Avg DK}
& \textbf{Avg $\Delta$} & \textbf{$t_{\mathrm{DK}}$ (s)} \\
\midrule
Qwen3-VL-4B-Instruct      & 4B  & 29.9 & 54.8 & $+$24.9 & 2.46 \\
Qwen3-VL-8B-Instruct      & 8B  & 36.3 & 48.3 & $+$12.0 & 3.89 \\
Gemma4-E4B-IT             & 8B  & 50.8 & 59.9 & $+$9.1  & 3.29 \\
Qwen3.5-9B                & 9B  & 64.6 & 80.9 & $+$16.3 & 4.92 \\
Step3-VL-10B              & 10B & 74.2 & 78.8 & $+$4.6  & 52.75 \\
Gemma4-12B-IT             & 12B & 54.9 & 67.4 & $+$12.6 & 4.43 \\
Qwen3.6-27B               & 27B & 71.4 & 86.1 & $+$14.7 & 20.77 \\
Qwen3-VL-30B-A3B-Instruct & 30B & 73.9 & 86.5 & $+$12.6 & 2.64 \\
\bottomrule
\end{tabular}%
}
\end{table}

The latency results further show that parameter count alone is a poor predictor of computational cost. 
Qwen3-VL-30B-A3B-Instruct achieves the highest average \sol accuracy while requiring only 2.64\,s per CAPTCHA, providing the strongest accuracy--latency trade-off among the evaluated backbones. 
By comparison, Qwen3.6-27B reaches a similar accuracy of 86.1\% but requires 20.77\,s, while Step3-VL-10B requires 52.75\,s despite achieving a lower final accuracy of 78.8\%. 
Overall, the results suggest that selecting an efficient backbone is more important than either parameter scaling or model category alone when deploying \sol.

\section{Conclusion}
\label{sec:conclusion}

In this work, we introduced \sol, an autonomous, self-improving agentic framework that redefines the approach to defeating dark web CAPTCHAs. 
By shifting from per-site model retraining and large parameter scaling toward dynamic, experience-driven tool orchestration, \sol enables lightweight, open-weight VLMs to reach high accuracy among deployable open-weight solvers.
%Our empirical evaluations reveal a critical capability plateau in raw VLM scaling: massive models (e.g., 122B) fail to meaningfully surpass smaller baselines on heavily obfuscated images and introduce severe latency bottlenecks that violate the operational constraints of Tor-based CTI crawling. 
%Conversely, 
Equipping a small 9B VLM with \sol yields a +36.8\% accuracy gain on the most heavily defended dataset, with the local agentic loop completing in 5.38s per CAPTCHA.
% Among the evaluated backbones, Qwen3-VL-30B-A3B-Instruct provides the best trade-off between accuracy and latency, achieving 86.5\% average accuracy across the four datasets while requiring only 2.64\,s of local inference time per CAPTCHA.
Furthermore, the dual-signal RAG architecture navigates the degeneration-recovery trade-off by learning to avoid  preprocessing steps that previously degraded similar CAPTCHAs.
In our evaluation, the agent's efficacy rests as much on restraint, knowing when to bypass preprocessing, as on active intervention. 

\textbf{Implications.} Beyond our specific tool, we see three implications; the last two are speculative. 
First, for CTI teams: in our case study, one workstation GPU, an open-weight model, and five classical image tools solved most CAPTCHAs without labelled data or per-market training. 
Practitioners can start from our results on backbone choice (the mixture-of-experts Qwen3-VL-30B-A3B gave the best accuracy--latency trade-off, Sec.~\ref{sec:inference-time}), on failure memory (Sec.~~\ref{sec:degen-recov} and \ref{sec:rag_ablation}), and on per-market warm-up (Sec.~\ref{subsec:feedback}). 
Second, server-rendered text CAPTCHAs, on the dark or clear web, may offer limited protection against low-cost local models; we expect operators to move to other challenge types or stricter rate limits, and collectors should plan for this. 
Third, learning from failures when not to act may benefit other agentic tool-use settings where a tool can degrade its input.

\section*{Acknowledgment}
 
This research/project is supported by the National Research Foundation, Singapore, and the Cyber Security Agency of Singapore under the National Cybersecurity R\&D Programme and the CyberSG R\&D Programme Office (Award CRPO-GC2-ASTAR-001). 
Any opinions, findings, conclusions, or recommendations expressed in these materials are those of the author(s) and do not reflect the views of the National Research Foundation, Singapore, the Cyber Security Agency of Singapore, or the CyberSG R\&D Programme Office.

% The preferred spelling of the word ``acknowledgment'' in America is without 
% an ``e'' after the ``g''. Avoid the stilted expression ``one of us (R. B. 
% G.) thanks $\ldots$''. Instead, try ``R. B. G. thanks$\ldots$''. Put sponsor 
% acknowledgments in the unnumbered footnote on the first page.

\bibliographystyle{IEEEtran}
\bibliography{ref}

\appendix

\subsection{Ethical Considerations}\label{sec:ethical}

This work studies CAPTCHA robustness in the context of dark web threat intelligence collection. While CAPTCHA-solving techniques can be misused to automate access to protected services (e.g., on the clear web), our objective is dark web analytics for the larger social good. 
We therefore release the research artifacts used in this study, including the annotated datasets, evaluation code, preprocessing tools, agentic pipeline, experiment configurations, and reproduction scripts.
To reduce unnecessary exposure of operational infrastructure, the released datasets contain no onion addresses, credentials, or other marketplace access information. 
% To reduce misuse potential, we do not release operational tooling, deployment code, or prompt templates. The paper provides sufficient methodological detail for scientific evaluation and reproducibility of the design, architecture, and findings, while withholding implementation artifacts that could directly facilitate CAPTCHA bypassing at scale.

% While CAPTCHA-solving techniques can be
% misused to automate access to protected services, our objective is to
% support reproducible research on dark web CTI collection. We therefore
% release the research artifacts used in this study, including the
% annotated datasets, evaluation code, preprocessing tools, agentic
% pipeline, experiment configurations, and reproduction scripts.
% To reduce unnecessary exposure of operational infrastructure, the
% released datasets contain no onion addresses, credentials, or other
% marketplace access information.

\subsection{Data Availability and Responsible Release}
\label{sec:opensource}

To support reproducibility and future research, we publicly release the complete source code, experiment configurations, reproduction scripts, and all four annotated CAPTCHA datasets at: \url{https://github.com/Binghui99/DarKnight}.
The released data contain no onion addresses or other marketplace access information. 
We also withhold operational solver assets and deployment tooling that could directly facilitate automated CAPTCHA bypass at scale.

% The CAPTCHA datasets were collected from real-world dark web marketplaces and could enable development or evaluation of automated CAPTCHA-solving systems. 
% To balance reproducibility with responsible disclosure, we do not publicly release the raw data. 
% Researchers with legitimate academic, defensive, or security research purposes may request access from the authors; requests are evaluated case-by-case and may require proof of institutional affiliation and agreement to use the data for research and educational purposes only.
% Under the same gated process, we can share methodological details sufficient to reproduce our design and experiments with vetted researchers, while withholding the raw data and operational artifacts that would facilitate misuse at scale.

\subsection{Inter-Annotator Agreement}
\label{app:annotation}
Table~\ref{tab:annotation_agreement} reports inter-annotator reliability for CAPTCHA transcription.
Three annotators participated in the annotation process, two of whom are Chinese-speaking, resulting in two annotators labelled 2,000 CAPTCHAs, while the third annotator skipped the 100+ Chinese CAPTCHAs from $\mathcal{D}_1$.
All agreement statistics were computed on independently produced transcriptions before any disagreement resolution.
We report exact-string agreement (ESA), which measures the proportion of independently produced transcription pairs that match exactly, and character error rate (CER), which quantifies character-level disagreement using edit distance.
For samples annotated by three annotators, the reported statistics aggregate agreement across all annotator pairs; for Chinese-script samples, agreement is evaluated between the two Chinese-speaking annotators.
After agreement measurement, disagreements were reviewed and resolved by consensus to establish the final ground-truth labels.

\begin{table}[htbp]
\centering
\caption{Inter-annotator agreement for CAPTCHA transcription. Agreement statistics were calculated before consensus resolution.}
\label{tab:annotation_agreement}
\resizebox{\columnwidth}{!}{
\begin{tabular}{lcccc}
\hline
\textbf{Dataset} &
\textbf{Samples} &
\textbf{ESA (\%)} &
\textbf{CER (\%)} &
\textbf{Resolution} \\
\hline
$\mathcal D_1$ & 500 & 90.4 & 2.90 & Consensus \\
$\mathcal D_2$ & 500 & 99.4 & 0.16 & Consensus \\
$\mathcal D_3$ & 500 & 96.4 & 0.70 & Consensus \\
$\mathcal D_4$ & 500 & 94.4 & 1.29 & Consensus \\
\hline
\end{tabular}
}
\end{table}

The results show high annotation agreement across all four datasets. 
D2 exhibits the highest reliability (99.4\% ESA and 0.16\% CER), consistent with its relatively clean visual structure, whereas $\mathcal{D}_1$ shows the largest disagreement (90.4\% ESA and 2.90\% CER), reflecting its greater visual and script complexity. 
Importantly, even for $\mathcal{D}_1$, character-level disagreement remains low, indicating that most transcription differences involve only a small number of characters.

% \subsection{Hardware and Software configuration}
% \label{app:hw_sw}
% All experiments were conducted on a single NVIDIA RTX Pro 6000 GPU (96 GB VRAM) running Debian 13 (trixie) with NVIDIA driver 610.57.04. 
% To simulate a resource-constrained production environment, we capped GPU memory at 30\% of its total capacity (~29 GB), using vLLM (v0.25.1) to enforce this limit. 
% All models were evaluated in their official, unmodified form: Qwen3-VL and Qwen3.5 (Qwen), Gemma4 (Google), and Step3-VL (StepFun). 
% No third-party quantization, fine-tuning, or other modifications were applied.

\begin{table*}[htbp]
\centering
\caption{Accuracy (\%) on dark web CAPTCHA datasets. \textbf{B} = baseline (raw VLM), \textbf{DK} = \sol (agentic pipeline). $\Delta$ = absolute change. Best agentic accuracy per dataset in \textbf{bold}. Models are ordered by size. HunyuanOCR is a pure OCR system that cannot be used as a Planner and therefore has no agentic results.}
\label{tab:main-results}
\small
\begin{tabular}{l c | c c c | c c c | c c c | c c c}
\toprule
& & \multicolumn{3}{c|}{\textbf{$\mathcal{D}_1$}} & \multicolumn{3}{c|}{\textbf{$\mathcal{D}_2$}} & \multicolumn{3}{c|}{\textbf{$\mathcal{D}_3$}} & \multicolumn{3}{c}{\textbf{$\mathcal{D}_4$}} \\
\textbf{Model} & \textbf{Params} & B & \textbf{DK} & $\Delta$ & B & \textbf{DK} & $\Delta$ & B & \textbf{DK} & $\Delta$ & B & \textbf{DK} & $\Delta$ \\
\midrule
HunyuanOCR         & 1B    & 12.4 & --            & --    & 61.4 & --            & --    & 45.7 & --            & --    & 90.0 & --            & --    \\
Gemma4-E4B-IT           & 8B  & 5.8  & 46.8                  & +41.0  & 71.4 & 65.8                  & $-$5.6 & 51.5 & 56.5                  & +5.0   & 74.4 & 70.4                  & $-$4.0 \\
Qwen3-VL-4B-Instruct    & 4B  & 28.0 & 55.6                  & +27.6  & 30.8 & 73.4                  & +42.6  & 26.8 & 35.3                  & +8.5   & 33.8 & 54.8                  & +21.0 \\
Qwen3-VL-8B-Instruct    & 8B  & 39.6 & 66.6                  & +27.0  & 43.4 & 49.0                  & +5.6   & 33.4 & 38.3                  & +4.9   & 28.8 & 39.4                  & +10.6 \\
Qwen3.5-9B              & 9B  & 36.6 & 73.4                  & +36.8  & 78.2 & 94.2                  & +16.0  & 58.1 & 63.4                  & +5.3   & 85.4 & 92.4                  & +7.0  \\
Step3-VL-10B            & 10B & 63.6 & 76.0                  & +12.4  & 98.6 & \textbf{100.0}        & +1.4   & 53.6 & 57.0                  & +3.4   & 80.8 & 82.0                  & +1.2  \\
Gemma4-12B-IT           & 12B & 22.4 & 46.0                  & +23.6  & 65.0 & 81.0                  & +16.0  & 43.0 & 47.7                  & +4.7   & 89.0 & \textbf{95.0}                  & +6.0  \\
Qwen3.6-27B             & 27B & 46.0 & 83.2                  & +37.2  & 85.2 & \textbf{100.0}        & +14.8  & 63.6 & 68.0                  & +4.4   & 90.6 & 93.0                  & +2.4  \\
Qwen3-VL-30B-A3B        & 30B & 42.4 & \textbf{80.6}         & +38.2  & 94.0 & \textbf{100.0}        & +6.0   & 66.7 & \textbf{72.3}         & +5.6   & 92.6 & 93.0                  & +0.4  \\
\midrule
\end{tabular}
\end{table*}

\subsection{Additional Baseline Comparison}
\label{app:additional_baselines}

To provide additional reference points against established OCR systems and frontier VLMs, we evaluate two locally hosted OCR baselines (PaddleOCR~\cite{paddleocr2020} and Tesseract~\cite{smith2007tesseract}) and four frontier Claude models on the ChangAn ($\mathcal{D}_1$) dataset. 
The Claude models are evaluated only as reference baselines and are not considered deployment candidates under the operational constraints discussed in Sec.~\ref{sec:intro}.

\begin{figure}[htbp]
    \centering
    \begin{subfigure}[t]{0.48\textwidth}
        \centering
        \includegraphics[width=\linewidth]{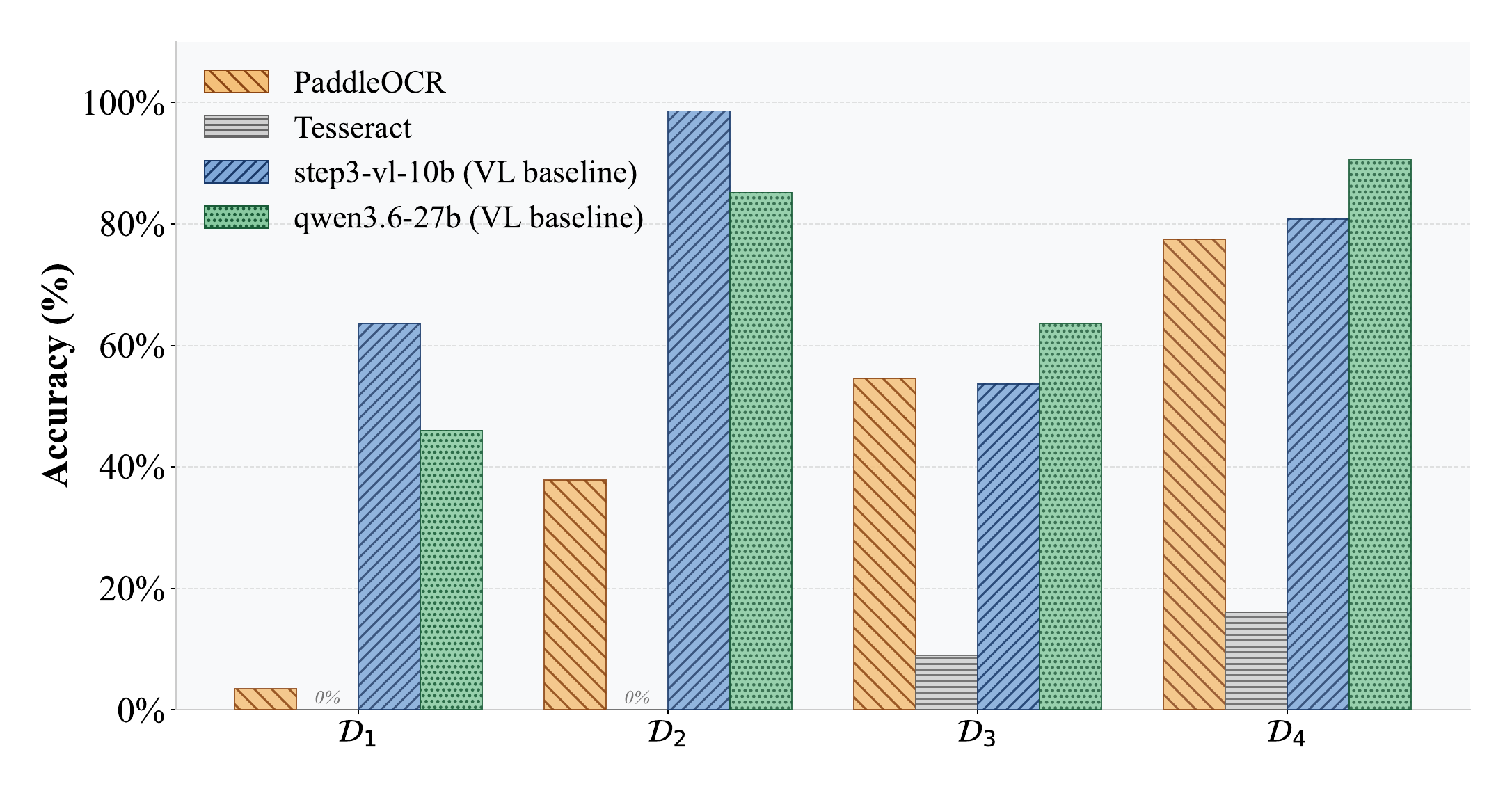}
        \caption{Classical OCR and VLM baselines across $\mathcal D_1-\mathcal D_4$.}
        \label{fig:ocr_baselines}
    \end{subfigure}
    \hfill
    \begin{subfigure}[t]{0.48\textwidth}
        \centering
        \includegraphics[width=\linewidth]{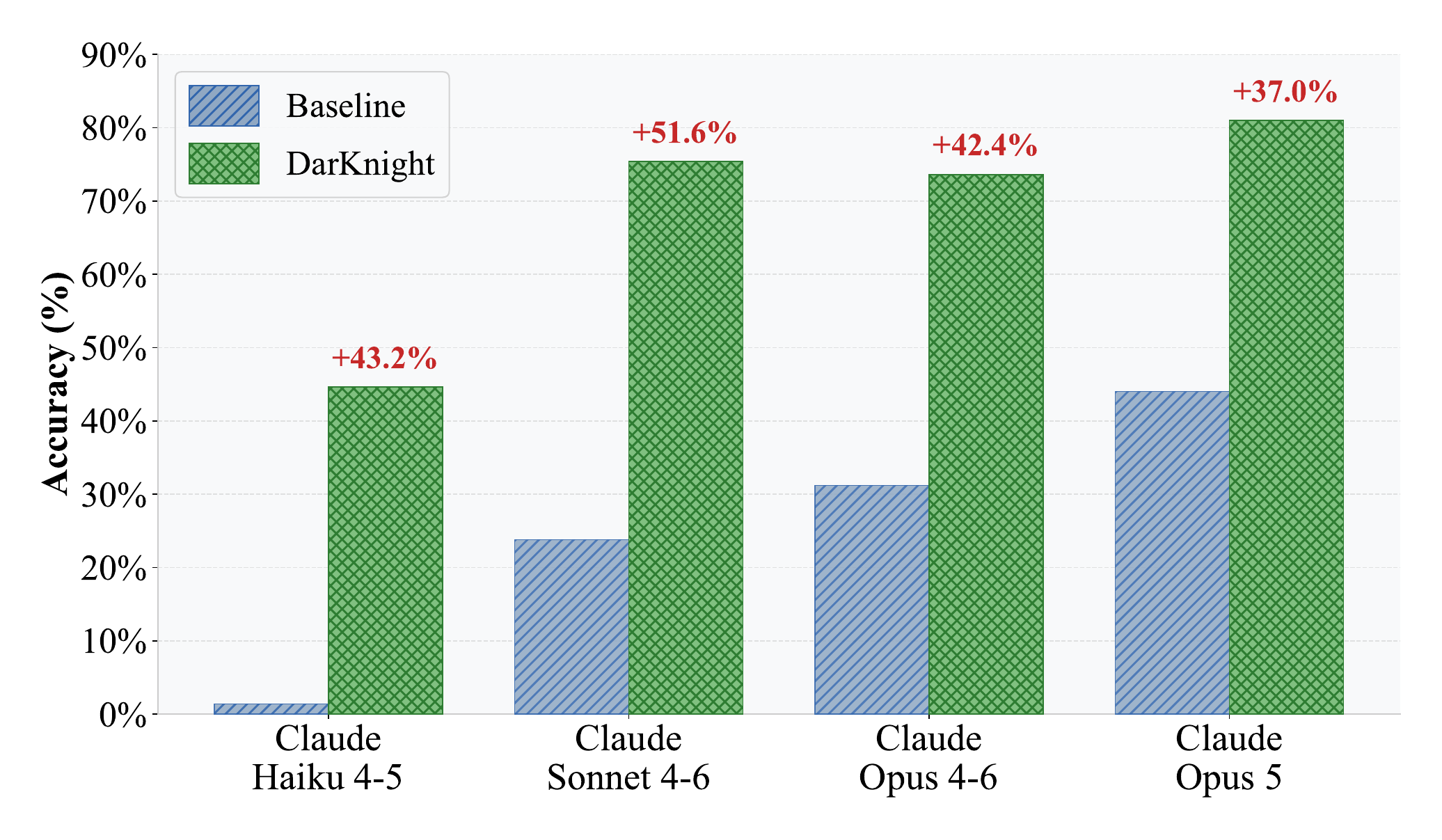}
        \caption{Frontier Claude models on $\mathcal D_1$, with and without \sol.}
        \label{fig:frontier_api}
    \end{subfigure}
    \caption{Comparison with other baselines.}
    \label{fig:additional_baselines}
\end{figure}

% \begin{table}[htbp]
% \centering
% \caption{Additional baseline comparison on the ChangAn ($\mathcal{D}_1$) dataset.
% Accuracy is exact-string match (\%).}
% \label{tab:additional_baselines}
% \resizebox{\columnwidth}{!}{
% \begin{tabular}{lccc}
% \hline
% \textbf{Method} &
% \textbf{Baseline} &
% \textbf{\sol} &
% \textbf{$\Delta$} \\
% \hline
% Tesseract          & 0.0  & --   & -- \\
% PaddleOCR          & 3.4  & --   & -- \\
% \hline
% Claude Haiku 4.5   & 1.4  & 44.6 & +43.2 \\
% Claude Sonnet 4.6  & 23.8 & 75.4 & +51.6 \\
% Claude Opus 4.6    & 31.2 & 73.6 & +42.4 \\
% Claude Opus 5      & 44.0 & 81.0 & +37.0 \\
% \hline
% \end{tabular}
% }
% \end{table}

\subsection{Sensitivity to RAG Memory Size}\label{app:rag_size}

% \hlt{We further evaluate the sensitivity of \sol to RAG memory size using Qwen3-VL-4B on $\mathcal{D}_1$ and $\mathcal{D}_2$.
% As shown in Table~\ref{tab:rag_size}, moderate memory sizes generally provide strong performance, while increasing the memory capacity does not consistently yield further improvements. 
% Among the evaluated settings, a memory size of 100 achieves the highest average accuracy across the two datasets and is therefore adopted as the default configuration. 
% These results suggest that \sol does not require an excessively large experience store to benefit from retrieval.}
We further evaluate the sensitivity of \sol to the per-slot RAG memory capacity using Qwen3-VL-4B on $\mathcal{D}_1$ and $\mathcal{D}_2$. 
Here, a RAG size of $N$ denotes a capacity of $N$ successful and $N$ failed records, for a maximum of $2N$ records in total. 
As shown in Table~\ref{tab:rag_size}, moderate memory sizes generally provide strong performance, while increasing the capacity does not consistently yield further improvements.
Among the evaluated settings, a per-slot capacity of 100 (up to 200 records in total) achieves the highest average accuracy across the two datasets and is therefore adopted as the default configuration.

\begin{table}[t]
\centering
\caption{Accuracy (\%) for varying RAG memory size.}
\label{tab:rag_size}
\begin{tabular}{c|cc}
\hline
RAG Size & $\mathcal{D}_1$ & $\mathcal{D}_2$ \\
\hline
No RAG & 52.8 & 51.0 \\
10  & 57.2 & 53.0 \\
20  & 56.8 & 73.2 \\
50  & 56.0 & 73.2 \\
80  & 57.2 & 64.8 \\
100 & 57.8 & 73.4 \\
150 & 56.8 & 73.2 \\
200 & 59.4 & 59.0 \\
\hline
\end{tabular}
\end{table}

The results in Fig.~\ref{fig:additional_baselines} show that conventional OCR systems perform poorly on ChangAn ($\mathcal{D}_1$), with PaddleOCR achieving 3.4\% and Tesseract 0\%.
The frontier Claude models provide substantially stronger raw performance, reaching up to 44.0\%. \sol further improves all four Claude models, with final accuracy reaching up to 81.0\%.

\subsection{API parameters for VLM calls}
\label{sec:apipara}
Tab.~\ref{tab:calls} specifies the exact hyperparameter configurations and inference payloads used for the two distinct VLM invocations (planning and execution) within the framework.
\begin{table}[htbp]
\centering
\caption{API parameters for the two VLM calls per sample.}
\label{tab:calls}
\resizebox{0.8\columnwidth}{!}{%
\begin{tabular}{lcc}
\toprule
\textbf{Parameter} & \textbf{visual\_analysis} \& \textbf{plan} & \textbf{run\_ocr} \\
\midrule
\texttt{max\_tokens} & 2048  & 1024 \\
\texttt{temperature} & 0.0 & 0.1 \\
Input & image+text+RAG & processed image+text \\
\bottomrule
\end{tabular}%
}
\end{table}

\subsection{Detailed results}
\label{sec:accuracytab}
This section provides the complete granular dataset for the evaluation discussed in Sec.~\ref{sec:results}. Table~\ref{tab:main-results} details the exact accuracy percentages and absolute changes ($\Delta$) for all evaluated VLMs across the four dark web datasets, serving as the numerical foundation for Fig.~\ref{fig:accuracy_charts}.

\subsection{Discussion}
\label{sec:discussion}

\noindent\textbf{Local CAPTCHA Solving for CTI.}
Historically, defeating specially designed dark web CAPTCHAs has required either labor-intensive per-site engineering or increasingly large vision models. 
\sol demonstrates that strong performance can instead be achieved by combining open-weight VLMs with adaptive, experience-driven tool orchestration.
In particular, our results show that model size alone does not determine either accuracy or inference cost; an efficient backbone paired with \sol can provide a favorable
accuracy--latency trade-off without relying on third-party VLM APIs.
This makes local CAPTCHA solving more accessible for automated CTI collection. % while preserving control over sensitive crawling data.

\noindent\textbf{The Necessity of Negative Memory in Visual Agents.}
A core insight from our evaluation is that agentic visual systems fail not just from an inability to act, but from an inability to restrain themselves.
When applied uniformly, aggressive preprocessing consistently degrades clean inputs. \sol overcomes this via its dual-signal RAG architecture. 
By treating historical failures as explicit negative constraints rather than discarding them, the Planner learns a conservative execution policy. 
This suggests that multimodal agents deployed in volatile security contexts may benefit from incorporating explicit failure-memory mechanisms to prevent self-induced degradation.

\noindent\textbf{Limitations and Adversarial Adaptation.}
While our evaluation uses real-world dark web CAPTCHA datasets, a live deployment faces dynamic adversarial conditions that are not fully captured by an offline study. 
First, marketplace operators may introduce targeted perturbations or entirely new CAPTCHA mechanisms designed to disrupt VLM-based solvers. 
Although \sol can adapt its preprocessing strategy through accumulated experience, its formal robustness against targeted adversarial examples remains an open question. 
Second, our latency measurements capture local computational overhead only. 
In a live deployment, end-to-end latency would additionally depend on variable Tor network delays and crawler request pacing, which we did not measure. 
Future work will integrate \sol into a live adaptive crawler to evaluate its long-term behavior under operator countermeasures and network variability.

\end{document}